%% file: 37404corr.tex
\documentclass[]{aa}
\usepackage{url}
\usepackage{natbib}               
\usepackage{graphicx}             
\usepackage{amssymb}              
\usepackage[varg]{txfonts}
\usepackage{relsize}

\include{definitions}

\begin{document}

\title{Magnetic twist profile inside magnetic clouds derived with a superposed epoch analysis}

\titlerunning{Twist distribution inside magnetic clouds}
\authorrunning{Lanabere et al.}

\author{Lanabere V.\inst{1}, Dasso S.\inst{1,2,3}, D\'emoulin P.\inst{4}, Janvier M.\inst{5}, Rodriguez L.\inst{6} \and Mas\'ias-Meza J.J.\inst{3}}
   \offprints{V. Lanabere}
\institute{
$^{1}$ Universidad de Buenos Aires, Facultad de Ciencias Exactas y Naturales, Departamento de Ciencias de la Atm\'osfera y los Oc\'eanos, 1428 Buenos Aires, Argentina, \email{vlanabere@at.fcen.uba.ar, sdasso@at.fcen.uba.ar}\\
$^{2}$ CONICET, Universidad de Buenos Aires, Instituto de Astronom\'\i a y F\'\i sica del Espacio, CC. 67, Suc. 28, 1428 Buenos Aires, Argentina, \email{sdasso@iafe.uba.ar} \\
$^{3}$ Universidad de Buenos Aires, Facultad de Ciencias Exactas y Naturales, Departamento de Física, 1428 Buenos Aires, Argentina, \email{dasso@df.uba.ar, masiasmj@df.uba.ar}\\
$^{4}$ LESIA, Observatoire de Paris, Universit\'e PSL, CNRS, Sorbonne Universit\'e, Univ. Paris Diderot, Sorbonne Paris Cit\'e, 5 place Jules Janssen, 92195 Meudon, France, \email{Pascal.Demoulin@obspm.fr}\\
$^{5}$ Universit\'e Paris-Saclay, CNRS, Institut d'Astrophysique Spatiale, 91405 Orsay, France \email{mjanvier@ias.u-psud.fr}\\
$^{6}$ Solar–Terrestrial Center of Excellence – SIDC, Royal Observatory of Belgium, Av. Circulaire 3, 1180 Brussels, Belgium\\ 
}

   \abstract  
   {Magnetic clouds (MCs) are large-scale interplanetary transient structures in the heliosphere that travel from the Sun into the interplanetary medium. The internal magnetic field lines inside the MCs are twisted, forming a flux rope (FR). This magnetic field structuring is determined by its initial solar configuration, by the processes involved during its eruption from the Sun, and by the dynamical evolution during its interaction with the ambient solar wind.}
   {One of the most important properties of the magnetic structure inside MCs is the twist of the field lines forming the FR (the number of turns per unit length).
   The detailed internal distribution of twist is under debate mainly because the magnetic field ({\bf B}) in MCs is observed only along the spacecraft trajectory, and thus it is necessary to complete observations with theoretical assumptions.
   Estimating the twist from the study of a single event is difficult because the field fluctuations significantly increase the noise of the observed {\bf B} time series and thus the bias of the deduced twist. 
    }
 {The superposed epoch applied to MCs has proven to be a powerful technique, permitting the extraction of their common features, and removing the peculiarity of individual cases.
 We apply a superposed epoch technique to analyse the magnetic components in the local FR frame of a significant sample of moderately asymmetric MCs observed at 1~au.
 }
{From the superposed profile of {\bf B} components in the FR frame, we determine the typical twist distribution in MCs. The twist is nearly uniform in the FR core (central half part), and it increases moderately, up to a factor two, towards the MC boundaries. This profile is close to the Lundquist field model limited to the FR core where the axial field component is above about one-third of its central value.
}
{} 
\maketitle


\section{Introduction} 
\label{sect_Introduction}
A coronal magnetic structure in equilibrium in the solar atmosphere can reach a global instability threshold when the magnetic stress becomes too high. 
Then, plasma can be ejected into the interplanetary medium and 
is observed as a coronal mass ejection (CME) by solar coronagraphs. 
These coronal remote white-light observations have shown that the distribution of mass is consistent with twisted structures in at least some CMEs that are well oriented along the line of sight. 
When CMEs are observed in the interplanetary medium, they are called interplanetary CMEs \citep[ICMEs, \eg][]{Wimmer-Schweingruber06S}. 

The link between CMEs and ICMEs has been well established for more than 30 years \citep[\eg][]{Sheeley85}. 
An important sub-set of ICMEs is known as magnetic clouds (MCs), a term introduced by \citet{Burlaga81}. 
An MC is characterised by in situ observations of an enhanced magnetic field strength with respect to ambient values, a smooth and high rotation of the magnetic field vector, and low proton temperature \citep[\eg][]{Burlaga81, Klein82, Burlaga95}.
The observed coherent rotation in MCs is interpreted as the passage of a large-scale twisted magnetic flux tube through the spacecraft. 

Magnetic twisted flux tubes, called flux ropes (FRs), are also present in several other systems in the heliosphere, such as the solar atmosphere, the solar wind, different locations of planetary magnetospheres, and ionospheres. 
FRs store and transport magnetic energy and helicity (H) \citep[\eg][]{Gosling90,Dasso09b,Nakwacki11,Demoulin16,Kilpua17}, and the distribution of twist is one major key for determining H.
Moreover, the twist distribution has consequences on the propagation of energetic particles inside MCs, in particular because the amount of twist modifies the field line length. \citet{Larson97} determined the length of the field lines in an MC as a function of the FR radius from both solar and in situ observations by tracking energetic particles.
\citet{Masson12} analysed the Earth-arrival time for energetic particles that are released from the Sun during relativistic solar particle events. 
They showed that this travel time for energetic particles is different within solar wind and within ICMEs.
Finally, the internal distribution of magnetic field in FRs, combined with the FR orientation, determines the temporal evolution of the magnetic field that is encountered by a planet. This means that field strength, twist, and the axis orientation of the FRs are key ingredients for the geoeffectiveness of MCs.

Magnetic clouds have being observed at different distances from the Sun several decades ago, and a diversity of models has been proposed to describe their internal magnetic structure.
One of the simplest models is an axially symmetric cylindrical magneto-static FR solution, with a relaxed linear force-free field, the so-called Lundquist model \citep{Lundquist50,Goldstein83}.
This model can describe the main features of the global field distribution for a significant number of observed MCs \citep[\eg][]{Burlaga82b, Burlaga95, Burlaga98, Lynch03, Dasso05, Lynch05, Dasso06, Lepping11}.  One main limitation of the model is that it frequently overestimates the axial field component near the flux-rope axis \citep[\eg][]{Gulisano05}. 

The distribution of the twist of field lines depends on the model. In the Lundquist model, the twist is nearly constant around the FR centre, and it increases outward as the axial field component decreases.  
Another model that is commonly used to describe MCs is the Gold-Hoyle model \citep[\eg][]{Farrugia99,Dasso03}; in this model, the twist is uniform. 

The mean number of turns is directly associated with the twist. For a small and hot flux rope, assuming the constant-twist GH model, \citet{Farrugia99} found a number of turns of $\sim$7au$^{-1}$.
For an MC that is observed by two spacecraft (STEREO and Wind) crossing different parts of the cloud and assuming a magnetic structure in a Grad-Shafranov equilibrium, \citet{Mostl09} found a small variation in the number of turns across the flux rope, with a mean value $\sim$2 au$^{-1}$.
Other models were also used to describe the FR in MCs \citep[\eg][]{Mulligan99,Hidalgo00,Hidalgo02,Cid02,Vandas03,Nieves09,Nieves16,Nieves18b}.  
In general, the distribution of the twist depends on the assumed model.

The vector magnetic field observed within MCs has a broad variety of profiles. This variety is narrower when the observations are rotated to the FR frame defined with the FR axis and the spacecraft trajectory. This variety further decreases when we take the sign of the magnetic helicity and only MCs that crossed down to their cores into account. Next, the common profile of the remnant variety is obtained by normalising and superposing the MC profiles. So far, this type of study has only been performed for scalar parameters such as the magnetic field intensity, proton density, and temperature \citep[\eg][]{Lepping03b,Yermolaev15,Masias-Meza16,Rodriguez16}. 
This method, known as superposed epoch analysis (SEA), avoids peculiarities of specific events, emphasising the common properties of the sample.

We here extend the work of \citet{Masias-Meza16} to the magnetic field components in order to derive the typical magnetic structure of FRs in MCs with a special focus on the distribution of the twist. In contrast to the case of scalar quantities, the FR orientation is needed, therefore we select events where the orientation is less biased, and we compare two different methods to obtain the orientation before making the superposition.  

In section \ref{sect_Data} we describe the sample of events we analysed and the quantities we studied.
We define the criteria we used to select cases, the methods with which we derived the FR orientation, and the parameter that quantifies the asymmetry of the magnetic field intensity. 
In Section \ref{sect_SEA}  we describe the implementation of the SEA. Then, we present the resulting superposed component profile of the field in the FR frame, as well as the fitting of these components with two different models. Finally, we present the typical twist profile in MCs that is obtained directly from SEA applied to the data of our sample. 
Finally, in Section \ref{sect_Conclusions} we present a summary and our conclusions. 

\section{Data and magnetic clouds }
\label{sect_Data}

\subsection{Magnetic cloud sample}
\label{sect_Data_MC}

\begin{figure*}[t!]           
\centering
\includegraphics[width=\textwidth, clip=]{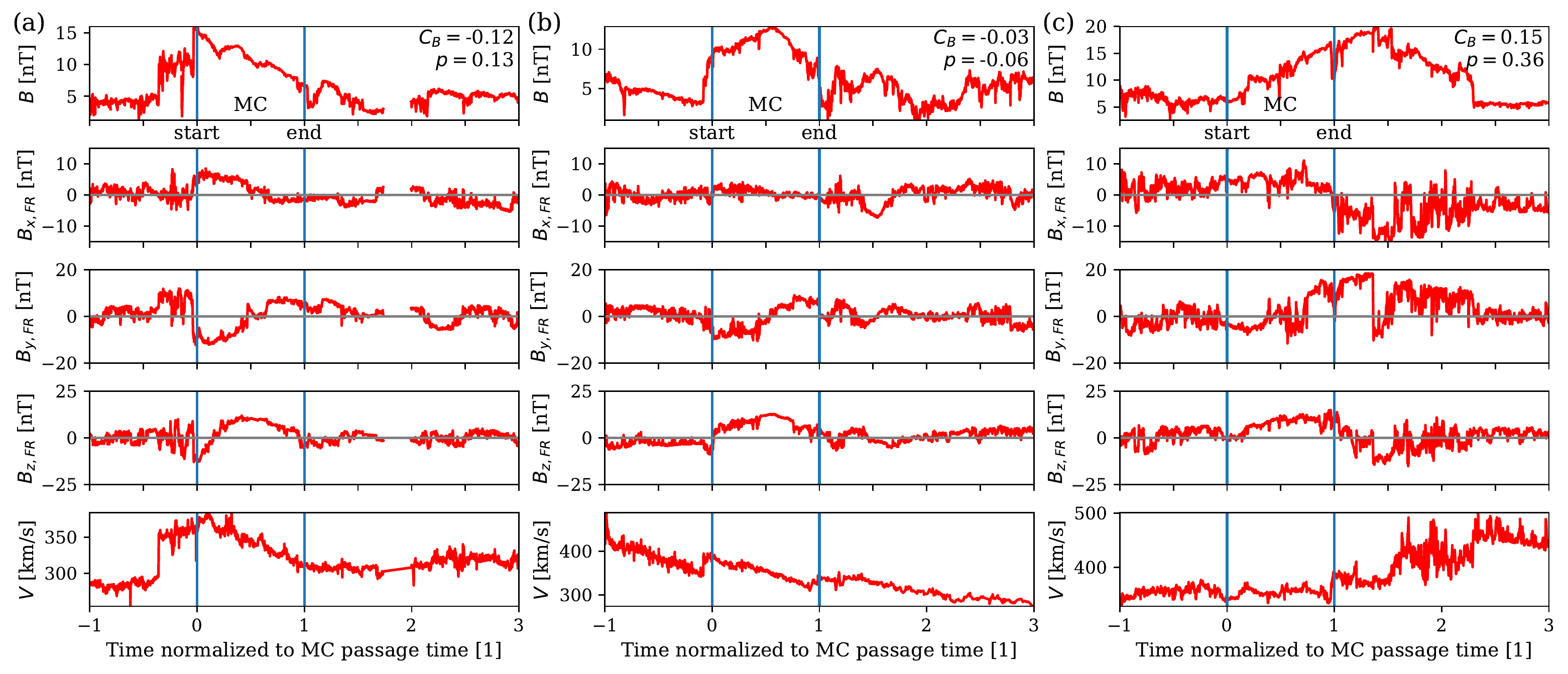}
\caption{Example of three MC profiles illustrating the variety of the observed asymmetric profiles of $B(t)$ for a comparable velocity magnitude. (a) 1 November 2012, (b) 4 March 1998, and (c) 3 September 2002.
  From top to bottom, the panels show the magnetic field magnitude ($B$), its components ($\BxFR$, $\ByFR$, $\BzFR$) in the FR frame (\fig{coordinate_system}), and the plasma velocity $V$. The time axis is normalised with the MC boundaries: $\tstart=0$ and $\tend=1$. 
}
\label{fig_events}
\end{figure*}

In situ measurements of the magnetic field and plasma properties in the interplanetary medium have been carried out by several spacecraft since the 1960s. In particular, the Wind spacecraft has been observing the interplanetary medium since 1994 at 1~au. For this study we used data from two instruments, the Magnetic Field Instrument (MFI) and the Solar Wind Experiment (SWE), with a temporal cadence of 60 s for (MFI) and 92 s for (SWE), which we downloaded from \url{https://cdaweb.sci.gsfc.nasa.gov/pub/data/wind/mfi/mfi_h0} and \url{https://cdaweb.sci.gsfc.nasa.gov/pub/data/wind/swe/swe_h1/} , respectively.

We examined the data for a set of MCs observed by Wind between 1995--2012 that were analysed by \citet{Lepping90, Lepping11}. The list of MCs can be found at \url{https://wind.gsfc.nasa.gov/mfi/mag_cloud_S1.html}. This table contains information about the start and end times of the passage of Wind through the MCs, the orientation of the flux rope axis, the closest-approach distance of the spacecraft from the FR axis, and other physical quantities derived from fitting the Lundquist magnetic field model \citep{Lepping90}.
The impact parameter is defined by normalising the closest-approach distance with the FR radius (see \ap{FRframe}). 
In this catalogue, each MC is classified into three different categories according to the fitting quality ($Q_0$), where $Q_0 = 1$ means that good fitting was obtained, $Q_0 = 2$ stands for fair quality, and $Q_0 = 3$ for poor quality, as defined in Appendix A of \citet{Lepping06}.

We restrict our analysis to the MCs from Lepping's catalogue that have $Q_0=1$ or $Q_0=2$, a sub-set that we call $\Qall$, so as to focus on the best-fit MCs. We analyse the magnetic field components ($\BxFR$, $\ByFR$, $\BzFR$) in the FR frame that is attached to the FR, with $\BzFR$ the axial component and ($\BxFR$, $\ByFR$) the components in the orthogonal directions (\ap{FRframe}). The FR axis orientation is provided by the angles $\tL$ and $\pL$. We also derive the FR axis orientation  by applying the minimum variance method (\sect{Data_Comparison_L_MV}).

\subsection{Magnetic cloud examples}
\label{sect_Data_Examples}

We show below three examples of MC data, rotated in the FR frame (\fig{coordinate_system}) using the orientation given by ($\tL$ and $\pL$), in order to present the type of data used.
  
Because the main purpose of this work is to derive the magnetic twist within MCs, we used the time series of the magnetic field strength and its components in the FR frame as the basic data input (top four rows of \fig{events}). In order to make a coherent superposition of different events regardless of their magnetic chirality and the sign of $p$, the sign of the $\BxFR$ and $\ByFR$ components were changed so as to have FRs with positive magnetic helicities and impact parameters (see \sect{SEA_method} for more information). The shape of the time profile of $B(t)$ within MCs is associated with the plasma bulk velocity \citep{Rodriguez16,Masias-Meza16}. Thus, we also analyse the plasma velocity in this study (bottom row of \fig{events}).

Because our aim in \sect{SEA} is to superpose a set of MCs with different time durations ($\Delta t$), we normalised the time so that the start time of all MCs corresponded to $t=0$ and the end time to $t=1$. We also extended the plot to $\Delta t$ before the start of the MC, and $2\, \Delta t$ after (larger after to include the long remaining trace of the MC passage).

We selected three MCs from the set $\Qall$ to illustrate the observed magnetic field components in the FR frame using Lepping's orientations ($\tL$ and $\pL$, \fig{events}). We selected MCs with comparable speeds, durations (about one day), and low impact parameters while sampling the variety of observed $B(t)$ profiles. More precisely, \fig{events}a shows an MC with a stronger field at the front, \fig{events}c shows stronger field at the rear, and \fig{events}b shows a case with a more symmetric profile, which according with \citet{Demoulin09b} is closer to the field expected in cylindrical FRs (\eg\ described by \eqs{B_FR}{B_Lundquist}).

The magnetic field components show the behaviour expected of typical FR models such as that of Lundquist, \eq{B_Lundquist}, or the one of Gold-Hoyle, \eq{B_GH}.
$\BxFR$ is typically weaker than other components since we select ed cases with a low impact parameter, implying that, except near the core where the azimuthal field is not significant, $\BxFR$ includes only a small fraction of the azimuthal field component (\eq{B_FR}).
$\ByFR$ has a sinusoidal-like shape, as is expected from the azimuthal field component of FR models, \eg\ with \eq{B_Lundquist} or \eq{B_GH} included in \eq{B_FR}.
$\BzFR$ is stronger in the FR core and decreases to $\BzFR \approx 0$ near the MC boundaries (fourth row in \fig{events}). The MC of panel (c) deviates the most from this typical pattern because it is overtaken by a fast magnetised solar wind stream (shown in the bottom row) that compresses the magnetic field (stronger B at the MC rear, \fig{events}c in the top row).

\subsection{Asymmetry parameter $\CB$}
\label{sect_Data_CB}

A sub-set of the observed MCs has a strong asymmetry in field strength $B(t)$ between the inbound and outbound direction (\ie\ between the trajectory of the spacecraft before and after it reaches the closest approach distance to the FR axis, respectively). The top row of \fig{events}a and c provides two examples. These profiles are different than the symmetric profile that is expected in cylindrical FR models. 
This asymmetry typically cannot be explained by the amount of FR expansion alone because it is crossed by the spacecraft (ageing effect) as this would require typically a much higher expansion rate than is observed \citep{Demoulin08,Demoulin18}. Most of this $B(t)$ asymmetry is instead the signature of a non-circular cross section of the FR \citep{Demoulin09b}.

Fitting a cylindrical model to asymmetric cases introduces a bias in the axis orientation.  
Another method that is frequently used to derive the orientation of FRs from in situ observations without assuming any specific FR model is the minimum variance (MV) method \citep[see \eg][]{Sonnerup67,Demoulin18}.
However, the MV method also provides an increasingly biased axis orientation with stronger field asymmetry. This bias is reduced, but is still present when the MV method is applied to the normalised magnetic field strength to unity \citep{Demoulin18}. Because the orientation of the FR axis is the key to rotating the field components in the FR frame, it is important to only retain cases with less bias in the axis orientation, that is, to filter out the MCs with a strong asymmetry. These are generally associated with spatial differences between the inbound and outbound branches of the FR \citep{Demoulin08}.

We quantify the $B(t)$ profile asymmetry with the centre of the magnetic field strength ($\CB$), which is analogous to the classical centre of mass (with time replacing spatial coordinates and $B$ replacing the mass density), as done by \citet{Janvier19}:
   \BE \label{eq_CB_definition}
    \CB = \frac{\int_{\tstart}^{\tend} \frac{t-\tc}{\tend-\tstart} \, B(t)\, \rmd t}
             { \int_{\tstart}^{\tend} B(t)\, \rmd t}
        = \frac{\int_{0}^{1} (t'-0.5) \, B(t')\, \rmd t'}{\Bmean}
   \EE
with the central time $\tc = (\tstart + \tend)/2$.  The normalisation of $t-\tc$ by the MC duration $(\tend-\tstart)$ at the denominator implies that $\CB$ is independent of the timescale
so that it is equivalent to apply it directly on the MC data or with a normalised time 
($\tstart=0$, $\tend =1$). 
For a symmetric $B$ profile around the time $\tc$, $\CB =0$.  $|\CB |$ increases with the magnetic field asymmetry, with $\CB$ negative when $B(t)$ is stronger before $\tc$ (\ie\ in the inbound), and positive when the field is more concentrated toward the MC rear (\ie\ in the outbound). $\CB$ is included in the interval $[-0.5,0.5]$ with the extreme values corresponding to an unrealistic field concentrated totally near the start or end boundaries, respectively.

\begin{figure}[t!]         
\centering
\includegraphics[width=0.5\textwidth, clip=]{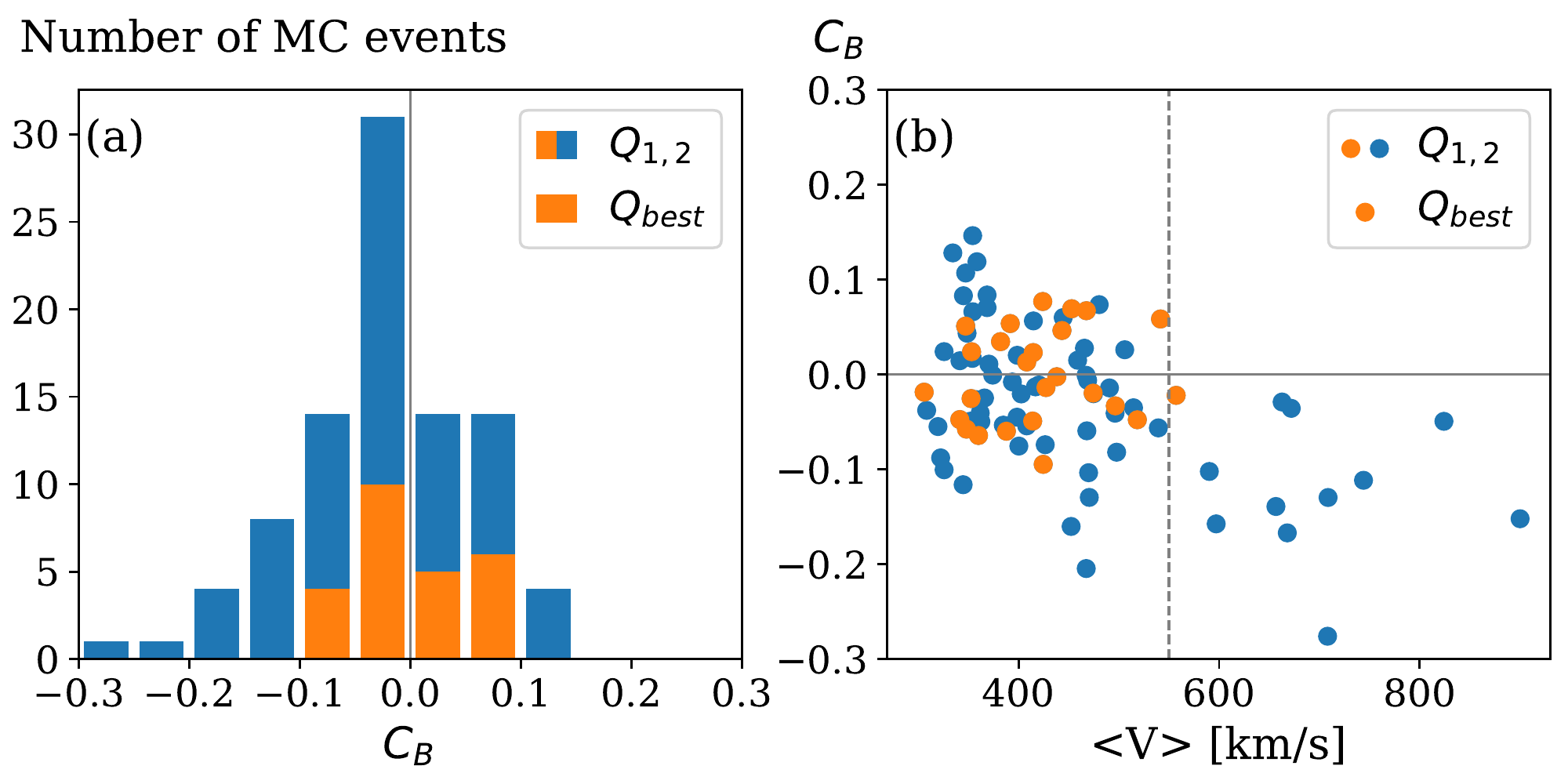}
\caption{(a) Histogram of the magnetic field asymmetry $\CB$ defined by \eq{CB_definition} and (b) scatter plot of $\CB$ vs. the mean MC velocity $\Vmean$. The group with $\Qall$ has 91 MCs and the group with $\Qbest$ has 25 MCs (included in $\Qall$). The dashed line in (b) corresponds to $\Vmean=550$\kms}
 \label{fig_histogram_CB}
\end{figure}

We computed $\CB$ using the trapezoidal rule for the discrete integration.  Histograms of $\CB$ are shown in \fig{histogram_CB}a with blue bars for the set $\Qall$. Almost $80\%$ of the MCs presents $|\CB|<0.1$.  The number of MC decreases sharply with higher $|\CB|$ values. 

The most asymmetric MCs of the examples shown in \fig{events} correspond to $\CB$ values between $\CB =-0.12$ and $\CB =0.15$, which are a priori surprising low values in view of their asymmetric $B(t)$ profiles in these extreme cases. Any $B(t)$ profile can be decomposed into the sum of a symmetric, $\Bs(t)$, and antisymmetric, $\Ba(t)$, functions around $t=\tc$. Then, the $\CB$ definition rewrites: 
   \BE \label{eq_CB_definition_b}
    \CB = \frac{\int_{\tstart}^{\tend} \frac{t-\tc}{\tend-\tstart} \, \Ba (t)\, \rmd t}
             { \int_{\tstart}^{\tend} \Bs(t)\, \rmd t}
        = \frac{\int_{0}^{1} (t'-0.5) \, \Ba(t')\, \rmd t'}{\Bmean}
   \EE 
The above small values of $|\CB|$ are partly due to the definition of $\CB$ and partly to the typical weak relative asymmetry present in many MCs (\ie\ $\Ba(t)$ is small compared with $\Bs(t)$).
The first row of \fig{events} provides a visual link between the shape of $B(t)$ and the value of $\CB$.

$\CB$ is correlated with the mean MC velocity (\fig{histogram_CB}b) and in a similar way with $B$ (not shown). In particular, all the cases travelling with $\Vmean$ larger than $550$\kms, or having $\Bmean$ larger than 23~nT, present a $B$ asymmetry with stronger field in the MC front (\ie\ $\CB<0$). This result is in agreement with the result reported by \citet{Masias-Meza16}, where the superposed epoch profile for fast MCs present a stronger $B$ field toward the MC front.

\begin{figure}[t!]        
\centering
\includegraphics[width=0.45\textwidth, clip=]{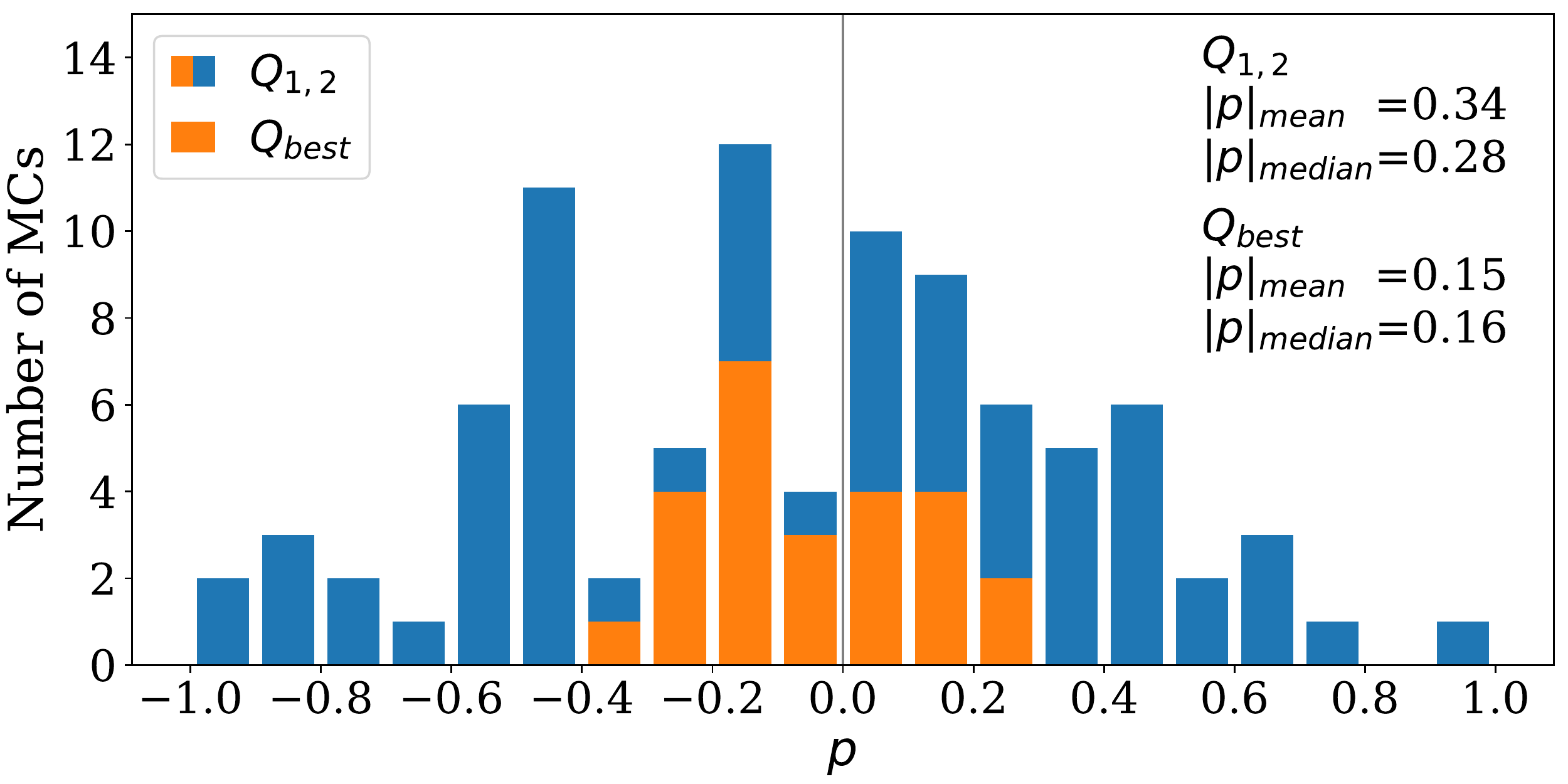}
\caption{Histograms of the impact parameter $p$ (ratio of the closest approach distance between the spacecraft trajectory and the FR axis). The group $\Qall$ of analysed MCs and its subset $\Qbest$ are shown.}
 \label{fig_histogram_p}
\end{figure}

\subsection{Magnetic could selection}
\label{sect_Data_MC_selection}

Our main aim is to derive the generic field components in well-behaved MCs (\ie\ smooth and coherent rotation $B$), and later on to derive the typical twist profile. This requires keeping mostly MCs that are well observed, that is, where the spacecraft trajectory corresponds to an approach distance to the FR axis that is not the closest possible position. Moreover, we needs to avoid MCs that are too perturbed (\ie\ strongly departing from the typical behaviour of MCs) during their evolution from the Sun to 1~au.   
For example, to exclude cases that interact strongly with another structure, such as another MC or an overtaking faster stream (\eg , \fig{events}c). The ideal cases would be isolated MCs, moving at the speed of the surrounding unstructured solar wind and crossed with a very small impact parameter. In practice, we need to relax these selection constraints in order to have a large enough number of cases.
The analysis of the field components in the FR frame requires first determining the  FR axis direction.  This was realised by independently applying the MV method and by performing a fit of the data with the Lundquist model. Both methods have an increasing bias in the deduced FR axis as $|p|$ and/or $|\CB|$ are larger, therefore we selected a lower range for both parameters. Furthermore, MCs with large negative/positive $\CB$ are extreme cases, which implies a strong compression at the front or rear. These cases are too rare to be analysed separately. Furthermore, as $|p|$ increases, a larger fraction of the azimuthal field is projected on $\BxFR$, and less on $\ByFR$ \citep[\eg][]{Demoulin18}, therefore it is important to limit our sample to low $|p|$ values. We therefore restrict the range of both $|\CB|$ and $|p|$ as much as possible below, with the constraint of keeping enough MCs to perform an SEA.

As $|\CB|$ increases, the associated FR is expected to have a cross-section shape that differs more strongly from a circular FR shape, and the determined FR axis orientation is expected to be more biased.
We therefore need to analyse only cases with low values of $|\CB|$. To retain enough cases for our SEA, we set the constraint to $|\CB|\leq 0.1$. 
This keeps 80\% of the MCs of quality $\Qall$ as the distribution of $\CB$ is concentrated around $\CB=0$ (\fig{histogram_CB}a), and this selection implies that asymmetric MCs such as the ones shown in \fig{events}ac are not kept.
 
For a given magnetic structure, the observed time series of the components $\BxFR$ and $\ByFR$ are expected to be significantly dependent on the impact parameter $p$ when the $|p|$ value is high (see \eq{B_FR}) while $|p|$ estimation has a significant error \citep[\eg ][]{Demoulin13}.
Moreover, a high $|p|$ value implies that only the external part of the FR is scanned, so this provides only a small amount of information on the FR. This implies typically an increasing bias on the deduced axis orientation with a larger $|p|$ value.  Then, we further select MCs with $|p|\leq 0.3$. This keeps 50 \% of the MCs as the distribution of $p$ for $\Qall$ is broad (\fig{histogram_p}).
 
To summarise, the set of MCs we used to apply the superposed epoch analysis contains cases with quality $\Qall$, with $|\CB|\leq0.1,$ and with $|p|\leq 0.3$. It contains 25 MCs that are the best cases observed within the original set of Lepping, so we call this set $\Qbest$.  The list of the selected MCs is shown in \tab{cases}.
We note that the distributions of $\CB$ for the selected cases, $\Qbest$ (in orange in \fig{histogram_CB}) are only weakly asymmetric so that biases introduced by positive and negative asymmetries are expected to partly cancel. All the MCs in this subset present a mean velocity inside the MC lower than $600$~\kms\ (\fig{histogram_CB}b).

\begin{figure}[t!]          
\centering
\includegraphics[width=0.5\textwidth, clip=]{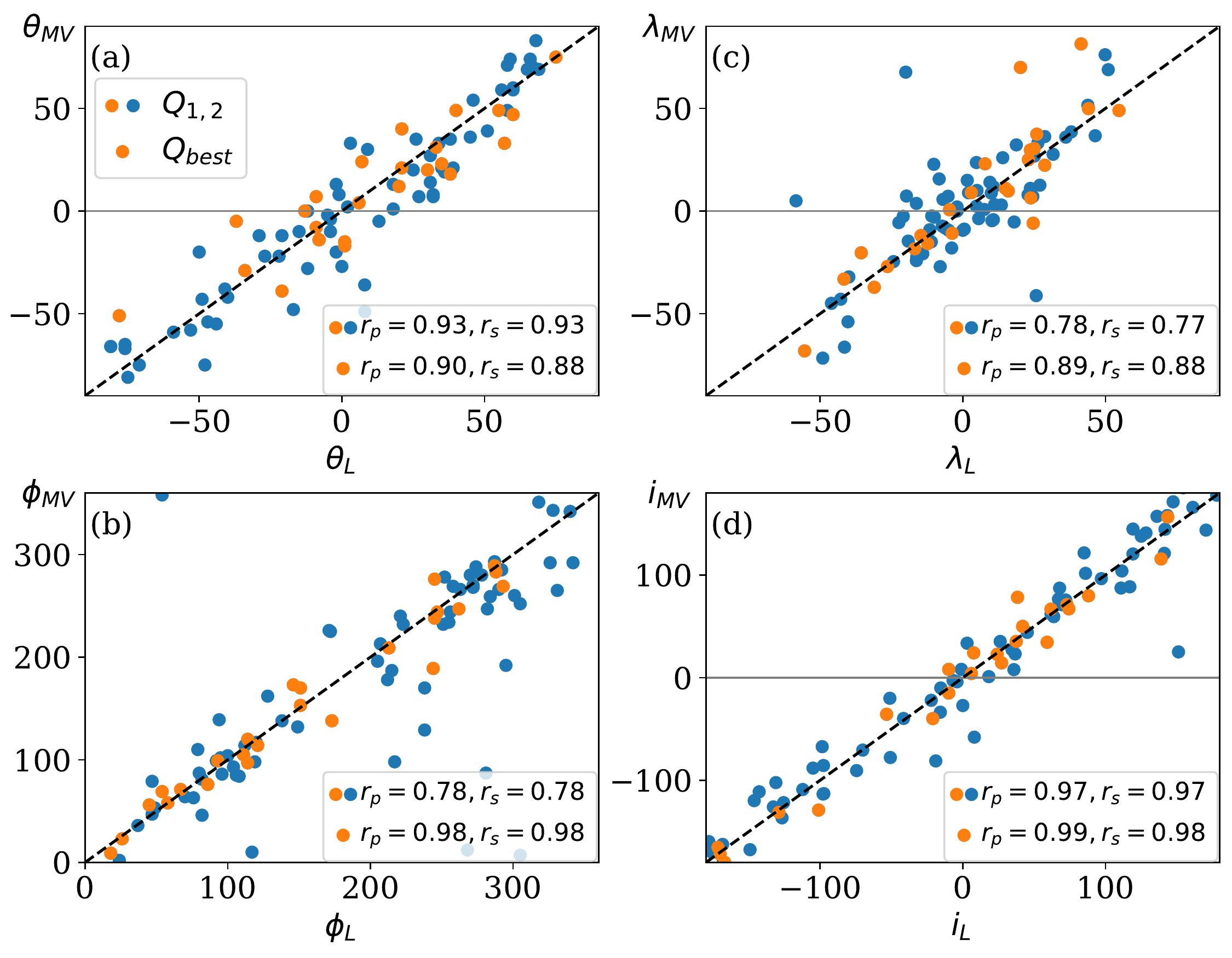}
\caption{Comparison of the four angles ($\theta$, $\phi$, $\lA$, and $\iA$) we used to define the FR axis orientation (\ap{FRframe}), which is computed with a Lundquist fit (in abscissa) and MV (in ordinate) methods. The Pearson, $\rp$, and Spearman, $\rs$, correlation coefficients are added in each subplot for the group $\Qall$ and its subset $\Qbest$ (see \sect{Data_MC_selection}).}
 \label{fig_comparison_orientation_L_vs_MV}
\end{figure}

\subsection{Comparison of Lepping and MV orientations}
\label{sect_Data_Comparison_L_MV}

The computed orientation of the FR axis is a key point of our analysis because it determines the field components in the FR frame.  Any method has intrinsic biases that at the present time are only partly understood. It is therefore worthwhile to compare the results of methods that are based on different hypotheses.

In contrast to the fit with a Lundquist model, \eq{B_Lundquist}, the MV method does not assume a specific FR model. Rather it is based on the expected different behaviours of the field components of a FR when expressed in the FR frame. As other methods, MV has biases which were recently studied in detail in \citet{Demoulin18}.  In particular the selection of the FR boundaries has a direct implication on the axis orientation bias.  Here,  we keep the same boundaries for the MV analysis than the ones used in the Lepping's list (at the expense of introducing more bias), since we look for a close comparison between both methods with the same input data.

The MV analysis associates the axis of the FR frame to extrema of the field component variance.
Specifically, $\BxFR$, $\ByFR$, and $\BzFR$ are associated with the directions of the lowest, highest, and intermediate variance, respectively.  However, each direction is not signed (an extremum corresponds to two opposite directions). Then, $\uzFR$, estimated by the MV, is set in the direction where $\BzFR>0$ in the FR centre, and $\uyFR$ direction is set closest to $\uzFR \times \ud$, with $\ud$ defined as the unit vector parallel to the trajectory of the spacecraft pointing to the Sun. Next, the $\uxFR$ direction is selected to have a right-handed FR frame.  Finally, this procedure defines an orthonormal frame, which is the MV estimation of the FR frame.  The FR orientation is summarised by the angles ($\tMV , \pMV$) or ($\lMV , \iMV$) as defined in \ap{FRframe}.
 
We applied the MV method to the same set of MCs defined as $\Qall$. The comparisons between the orientations found by Lepping ($\tL$, $\pL$) ($\lL$, $\iL$) and our MV ($\tMV$, $\pMV$) ($\lMV$, $\iMV$) are shown in \fig{comparison_orientation_L_vs_MV}.
We first describe the results with the MCs of the set $\Qall$ that do not belong to $\Qbest$ (blue colour). Some events present huge differences in $\phi$ between the two methods, greater than $60 \degree$ in absolute value. This arises because the polar axis of the selected spherical coordinate system lies in a region where some of the FR axes are located. For these FR axes, a small error on the axis orientation implies a large change of $\phi$ (\ap{FRframe}).  

As expected (\ap{FRframe}), no such large differences are found with the coordinates ($\lA, \iA$) because the differences are mostly all below $60 \degree$ in absolute value  (\fig{comparison_orientation_L_vs_MV}c,d). More precisely, a difference of the order of $60 \degree$ between Lepping and MV methods is only present for three MCs for $\lA$ and for one MC for $\iA$. 
Moreover, the correlation coefficients (Pearson and Spearman) of $\Qall$ are about 0.78 and 0.97 for $\lA$ and $\iA$, respectively.  As expected, the best determined angle is $i$ since it is associated to the nearly anti-symmetric component, $\ByFR$ which is well separated from the more symmetric components $\BxFR$ and $\BzFR$ \citep{Demoulin18}.  The partial mixing of $\BxFR$ and $\BzFR$ is different for Lepping and MV methods which introduces more dispersion of the $\lL$ and $\lMV$ values (\fig{comparison_orientation_L_vs_MV}c).   

The results with the set $\Qbest$ (orange) have even fewer cases with large absolute differences, as expected because the most difficult cases are removed with the criteria on $\CB$ and $p$ (\fig{comparison_orientation_L_vs_MV}c,d). This implies that the FR frames found with Lepping and MV methods are similarly oriented.  

\section{Generic magnetic field components}
\label{sect_SEA}

\subsection{Procedure of the superposed epoch analysis}
\label{sect_SEA_method}

The main aim of the SEA is to obtain a typical profile by taking a sample of individual profiles. Then, the results of the SEA emphasise the common characteristics of a data set, but minimise the peculiarities of individual events. This assumes that within the defined event boundaries,
the various events have a common behaviour for the analysed quantities so that the SEA result is physically meaningful. This is the purpose of \sect{Data_MC_selection}, where a careful selection of events was presented.

In order to provide meaningful results with vector data, that is, signed quantities, the physical properties are required to add up. This requires plotting the data in the FR frame where each magnetic field component has a precise meaning.   
The FR frame was defined so that $\BzFR>0$ in the FR core, then a direct SEA can be applied on $\BzFR$ on any MC sample. In contrast, $\BxFR$ and $\ByFR$ components are projections of the azimuthal component $\Bt$, and their signs depend on the signs of the FR helicity, $H$, and of the impact parameter, $p$.
Because $H$ is about equally distributed between positive and negative values \citep[\eg][]{Demoulin16b}, a direct SEA of $\BxFR$ and $\ByFR$ components would mostly result in the cancellation of the signal. We therefore set all MC cases to positive helicity.
A similar cancellation of the signal would further occur for $\BxFR$ which depends also on the sign of $p$ (\eq{B_FR}). $p$ is also almost equally distributed between positive and negative values (\fig{histogram_p}). 
In summary, we changed $\BxFR$ to $\sgn(p~H)~\BxFR$ and $\ByFR$ to $\sgn(H)~\ByFR$ to perform a coherent SEA on FRs with positive helicity and positive impact parameter. This implies that $\BxFR$ is dominantly positive, and $\ByFR$ changes from negative to positive values in all included MCs, as shown in the examples of \fig{events}.

The selected data extend on a time duration equal to the duration of the MC ($\Delta t$) before the front of the MC, and $2\Delta t$ after the end of the MC. This choice was made to analyse the solar wind surrounding MCs. In particular, we kept a longer time interval after the MC to analyse the trace of the MC perturbations after its passage.

We normalised the time of each MC to unity in such a way that it starts at $\tstart=0$ and ends at $\tend=1$, as in \fig{events}; this definition implies that negative time values correspond to data points before the MC arrival.
The SEA requires that the profile for each MC has the same number of data points. Because MCs do not present the same duration, the associated data differ in the number of data points. 
In order to obtain the same number of data points in every MC profile, we defined a grid of equally spaced bins, so that there are 50 bins inside each MC and 200 points in total for the full SEA profile. All the data points that fill in each SEA bin are averaged (mean value) to a single value.

This procedure was applied to all $\Qbest$ MCs present in \tab{cases}.
Finally, we computed the mean, median, and standard deviation at each SEA bin point. This procedure was used for the magnetic field intensity, magnetic field components in the FR frame, and for the solar wind speed.

\begin{figure}[t!]            
\centering
\includegraphics[width=0.5\textwidth, clip=]{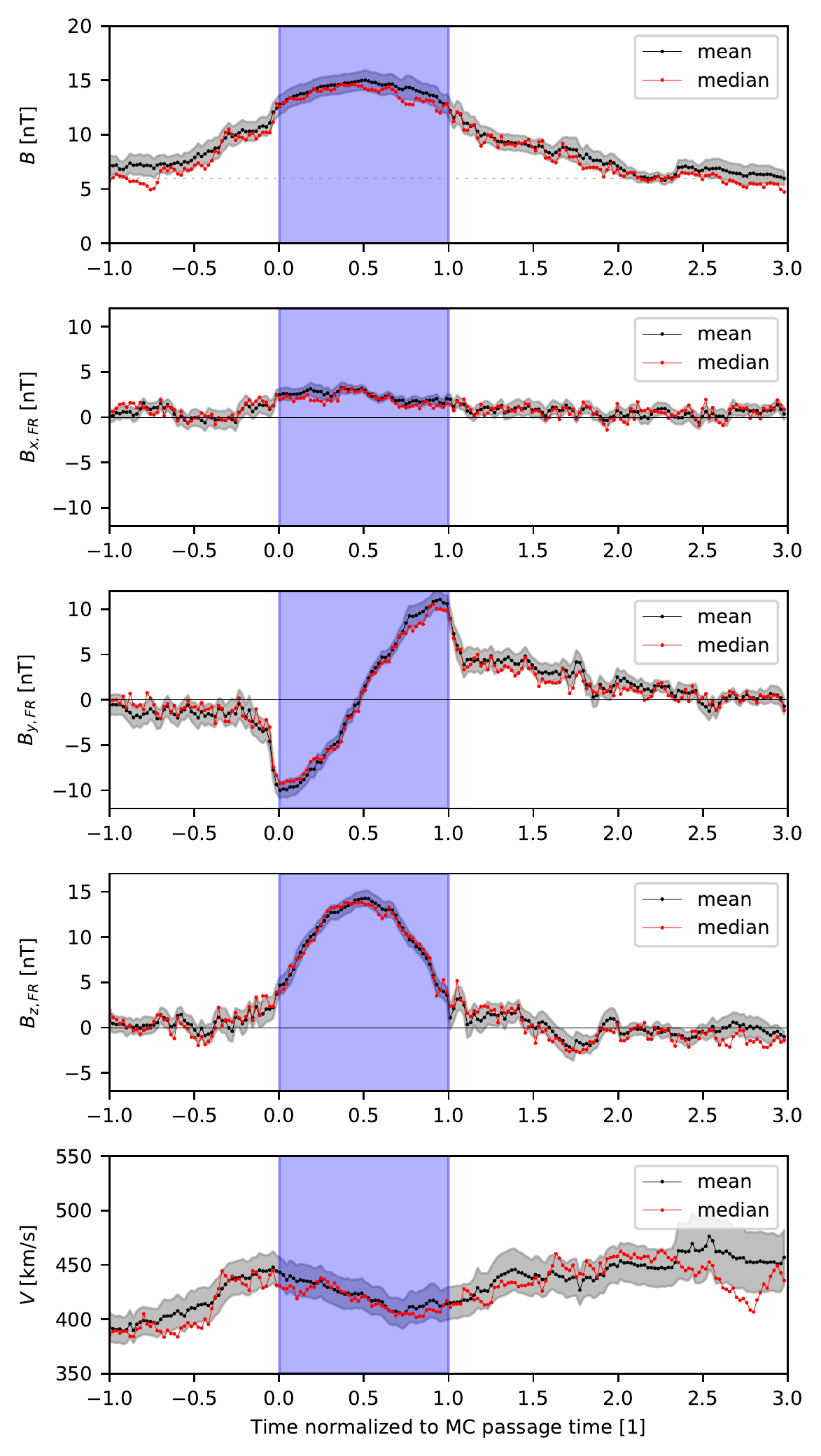}
\caption{Superposed epoch profile of the magnetic field (B and components) in FR frame with the Lepping orientation, and the plasma velocity (at the bottom). The profiles were computed using the mean (black) or the median (red) of the selected MCs ($\Qbest$ sample) in each temporal bin. The grey bands show the standard deviation of the mean. The MC region is shown with a blue background. The dashed line in the upper plot corresponds to $B=6$~nT associated with the ambient solar wind. The time is normalised by the MC duration ($\tstart=0$, $\tend=1$). 
   }
 \label{fig_superposed_L}
\end{figure}

\begin{figure}[t!]            
\centering
\includegraphics[width=0.5\textwidth, clip=]{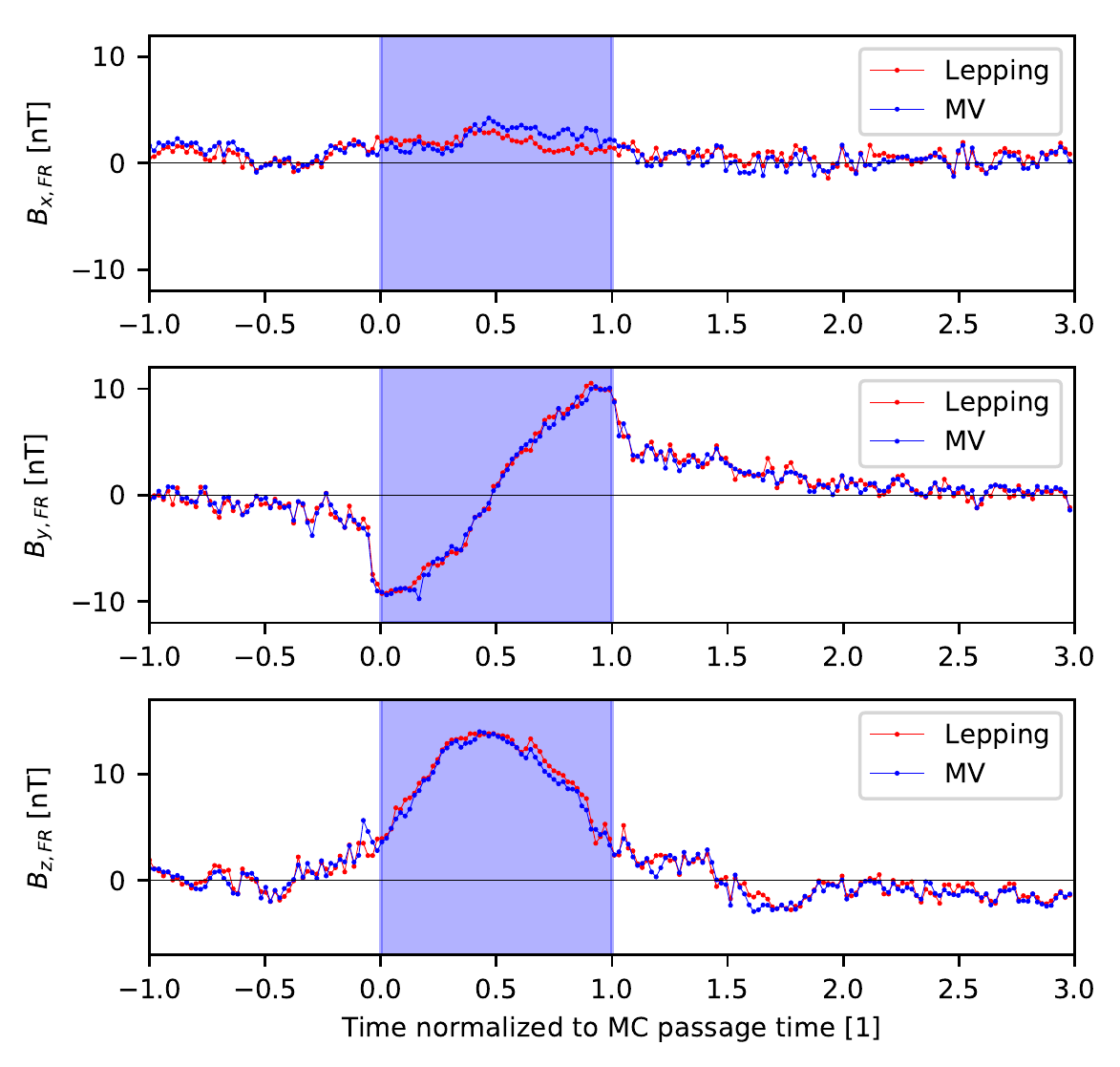}
\caption{Superposed epoch of the median profile of B components in the FR frame with Lepping (red) and MV (blue) orientations (with other drawing conventions as in \fig{superposed_L}).}
 \label{fig_superposed_L_vs_MV}
\end{figure}

\subsection{Superposed profiles in MCs}
\label{sect_SEA_profiles}

The SEA profiles are shown in \fig{superposed_L} with the Lepping FR orientation for the set $\Qbest$.  The SEA means are shown with black lines together with the associated errors of the means (grey bands), and the median values (red lines). Differences between the mean and median values are typically under the error bar.

As expected, the profiles for the different quantities shown in \fig{superposed_L} are significantly smoother than the time profile of the same quantities for individual cases (\eg\ compared with profiles shown in \fig{events}). This is a consequence of the superposed epoch technique, which can keep the common features but removes the peculiarities of individual events. While individual cases of the analysed sample ($Q_{best}$) can have an asymmetric profile of $B(t)$, this asymmetry is mostly removed in the superposed epoch, mainly because the distribution of $\CB$ is roughly symmetric (\ie\ about half of cases have the peak of $B$ before or after the FR central time, \fig{histogram_CB}). Moreover, the magnitude of the magnetic field inside the MC is about twice the magnitude of the ambient solar wind shown as a dashed line with $B=6$~nT.

$\BxFR (t)$ is positive because both the impact parameter and helicity were converted to be positive (see \sect{SEA_method}). Still, $\BxFR$ is small compared to $B$, as expected due to our selection of cases for the analysed set of $Q_{best}$ (low $|p|$ values, see \fig{histogram_p}).

The profile of $\ByFR (t)$ is nearly anti-symmetric around $t=0.5$, which is consistent with a closed flux rope that has the same amount of magnetic flux in the in- and outbound branches \citep[\eg][]{Dasso06}. $|\ByFR (t)|$ is maximum near both boundaries. Finally, there is a sharp decrease in $|\ByFR (t)|$ across the MC boundaries to about one-third and one-half its value inside at the in- and outbound boundary, respectively.  This clearly defines the FR borders, which for this analysis were defined for each MC by the catalogue of Lepping (see \sect{Data_MC}).

The profile of $\BzFR (t)$ is nearly symmetric, with its maximum value near the centre of the MC time interval. The values of $\BzFR$ at the two MC boundaries are low but positive with $\BzFR \sim 2.5$~nT. This corresponds to almost 20\% of the $\BzFR$ central value.  

$V(t)$ has a linear profile within most of the MC time interval. This is a classical signature of expansion \citep[\eg][]{Demoulin08,Gulisano10}. However, this expansion has only a weak effect on the measured magnetic field components because the profiles are almost symmetric or anti-symmetric around $t=0.5$. In other words, there is a small expansion of the FR during the spacecraft crossing, as expected with the typical observed expansion rates and FR sizes \citep{Demoulin08}.

Finally, very similar results are obtained with the Lepping and MV orientation of the FR for all field components (\fig{superposed_L_vs_MV}). This is a consequence of the good correspondances found between the axis angles ($\lL$, $\iL$) and ($\lMV$, $\iMV$) in \fig{comparison_orientation_L_vs_MV}. Even more, the fluctuations around the identity lines introduce almost symmetric differences for individual MCs which are mostly washed out with SEA.  
This is an important result as the biases of the Lundquist fit and MV are different, and moreover the MV method does not include a FR model, so it is not biased towards a specific magnetic structure model. It shows the strength of the SEA which minimises, by averaging, the differences in the FR axis orientations for individual cases. 

\subsection{Superposed profiles around MCs}
\label{sect_SEA_around_MCs}

The magnetic profile $B(t)$ before the FR boundary increases progressively from $B \approx 7$~nT in the solar wind to about $B \approx 11$~nT before the front boundary (\fig{superposed_L}). In particular, no trace of the front shock remains. This is due to a mixture of sheath sizes as a specific SEA of the sheaths was not performed here \citep[see][ for such SEA of sheaths]{Masias-Meza16, Janvier19}.

The magnetic field components before and after the front boundaries have the same sign. This occurs in the sheath during a time interval of about 20\% of the MC duration (\fig{superposed_L}). This is unlikely to be due to a mixture of the sheath and MC fields because the front boundary of MCs is typically well defined. We interpret this result as the consequence of the accumulation of magnetic field lines with similar direction in front of MCs, while those with significantly different direction are forced to reconnect as they are pushed against each others. This has an implication on the sheath size combined with erosion of the FRs as reconnection allows to evacuate more easily the accumulated plasma and magnetic field on the FR sides.

$B(t)$ stays above the SW background level after the MC for a duration slightly longer than the MC. This is coherent with the results of \citet{Masias-Meza16} (their Fig. 4) considering only the slow and medium velocity cases as we have done implicitly here with the selected criteria (\fig{histogram_CB}b). 
This is also coherent, while with a shorter timescale, with the results of \citet{Temmer17},  who found recovering times of 2--5 days after ICMEs. 
Next, the presence of an enhanced field after the MC is more important for $\ByFR$. As in previous studies, we interpret this result as the presence of a back region which is the remnant of the magnetic field present close to the Sun, at the periphery of the FR but which was reconnected on the front side with the encountered magnetic field \citep{Dasso06,Dasso07,Ruffenach15}. Then, starting from the MC centre ($\ByFR=0$) and including the extended region of $\ByFR>0$ after the MC interval provides an estimation of the typical amount of flux involved in the FR launched from the Sun. Finally, an enhanced $\BzFR$ is also present after the MC, but with both weaker value and duration as this component can be more easily transported away by Alfvén waves than $\ByFR$ component (whose total flux remains behind the MC in the absence of another reconnection).
 
Away from the FR boundaries, the magnetic field components are expected to outline a Parker spiral if they were superposed within a frame that is defined radial and orthoradial to the Sun and if the field direction was changed to have always the same flux sign away from the Sun.  None of these conditions is satisfied with the above SEA devoted to MCs. 
In particular, through the mix of FR orientations (\fig{comparison_orientation_L_vs_MV}), the components of the magnetic field away from the FR boundaries are cancelling to values near to zero as expected.
On the front side this occurs at a similar duration from the MC front boundary for all components.  This duration, about $0.2$ of the MC duration, is characteristic of the superposed sheaths.
On the MC rear side, the components vanish after different durations.  They range, in units of MC duration, from $\approx 0.1$ for $\BxFR$, to $\approx 0.4$ for $\BzFR$, to $\approx 1.2$  for $\ByFR$.

Finally, a clear mean overtaking faster stream is present at the rear in the SEA of velocity (\fig{superposed_L} bottom). This stream affects the MC rear over a duration of about one- third of the MC. Indeed, the majority of MCs, especially the slow ones, have an overtaking flow behind \citep{Rodriguez16}. This is also comparable to the SEA of slow and medium speed MCs \citep{Masias-Meza16}, which are the typical cases within $\Qbest$ sample as shown in \fig{histogram_CB}.

\begin{figure}[t!]           
\centering
 \includegraphics[width=0.5\textwidth, clip=]{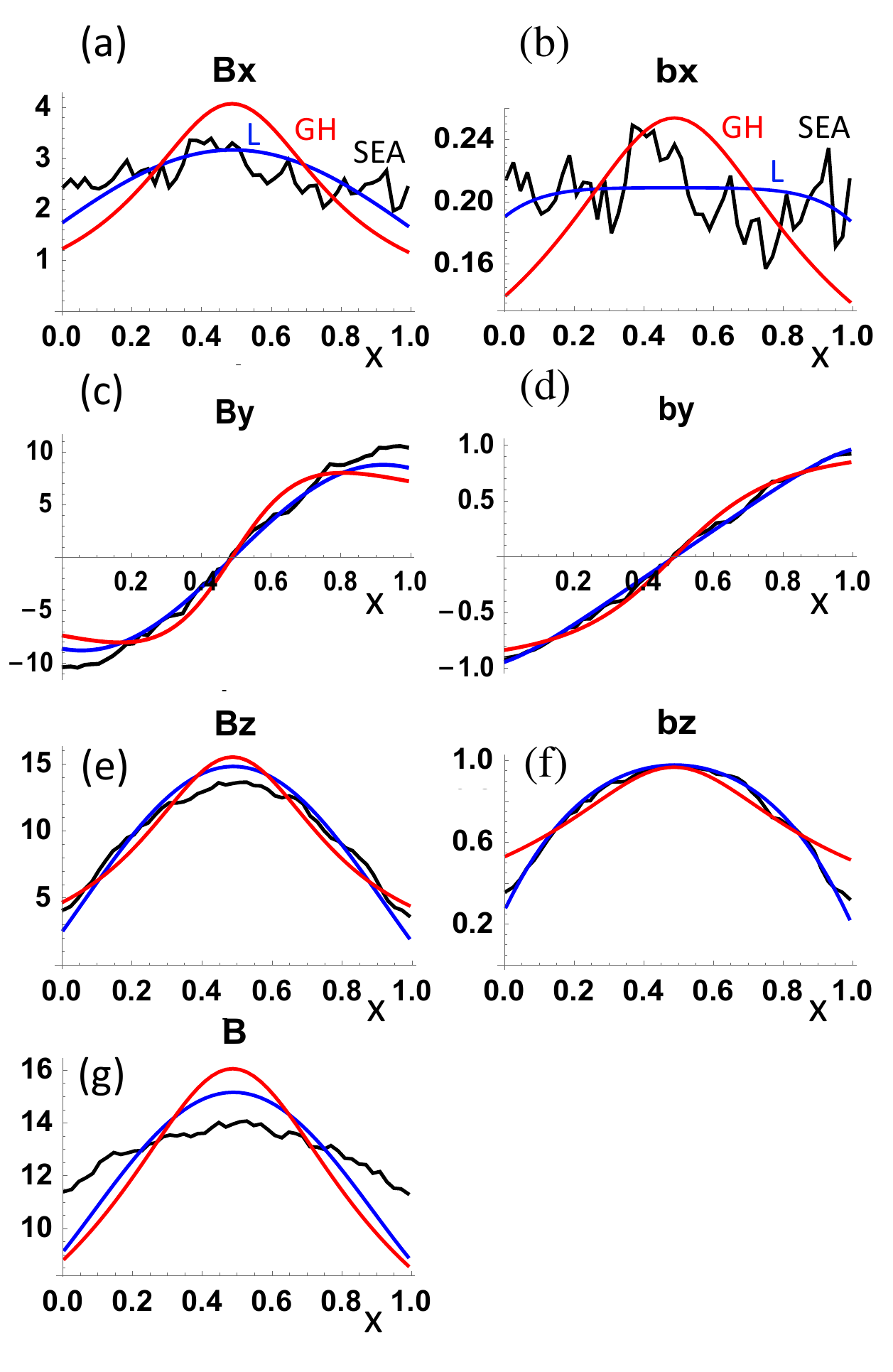}
\caption{Results of the mean SEA profile for $B$ and its components in the FR frame derived with MV (black lines) are fitted with the Lundquist (\eq{B_Lundquist}) and Gold-Hoyle (\eq{B_GH}) models (blue and red curves, respectively). (a, c, e, g) field in nT, (b, d, f) field components with $B$ normalised to unity.}    
 \label{fig_fit_B_SEA}
\end{figure}
 
\subsection{Fitting the SEA profile with models}
   \label{sect_SEA_Fitting}

We fit the SEA profiles with two standard FR models: the Lundquist (L, \eq{B_Lundquist}) and the Gold-Hoyle (GH, \eq{B_GH}). The normalised time of SEA profiles is simply converted in to a spatial $x$ coordinate by assuming a constant velocity. This is justified by the symmetry of the SEA profiles which implies a negligible effect of the FR expansion on the SEA profiles. Then, the $x$ coordinate is equivalent to the normalised time.   

In the fitting procedure the models are described with free parameters.    
We allowed the FR model to be rescaled in size, then the normalised FR radius, $r_0$, is one parameter. Two other parameters describe the location of the FR axis, $\xo$ on the spacecraft trajectory and $\yp$ in the direction both orthogonal to the trajectory and to the FR axis. Normalising $\yp$ by the determined FR radius provides the impact parameter $p$. The rescaling in field strength is insured by the parameter $\Bo$ , which defines the field strength on the FR axis. Because the SEA results are in the FR frame, we do not introduce additional parameters to describe the FR axis orientation.
In summary, four free parameters ($r_0,\xo,p,\Bo$) are involved in the fit of both models.

The free parameters were computed by minimising the sum of the square of the differences between SEA and model profiles, that is, we performed a classical $\chi^2$ minimisation.  The quality of the fit is characterised by the square root of the reduced $\chi^2$ , defined as $\chiB=\sqrt{\chi^2/(3N-n)}$, $N=50$ is the number of data points, and $n=4$ is the number of free parameters. We also performed another fit with the observation vectors and model normalised to unity, therefore we fit only the vector directions. This type of fit is classically used in fitting a model to data of individual MCs \citep[\eg][]{Lepping90, Dasso03}. In this case $\Bo = 1$ and  $n=3$.  The quality of the fit is characterised as above by $\chib=\sqrt{\chi_{\rm b}^2/(3N-3),}$ where $\chi_{\rm b}^2$ is computed with the normalised vectors.

Examples of fitting results are shown in \fig{fit_B_SEA} with the SEA computed with the mean in each bin (the result is comparable when using the median shown in \fig{superposed_L_vs_MV}). We select the SEA profiles associated with the FR frame computed with the MV, as it is not explicitly biased by a specific model. The Lundquist model shows the best fit both in field strength and for all components. This is confirmed by a lower $\chiB$ value (\tab{fit}). The Gold-Hoyle model has a magnetic field ,which is too peaked in the FR core (\fig{fit_B_SEA}a,e,g), and which is too weak for $\BxFR$ and $\ByFR$ near the FR boundaries. The same conclusion is reached by using the Lepping frame or the median to build the SEA as all $\chiB$ values are comparable with a given model (\tab{fit}). This better fit of the data with the Lundquist model compare to the Gold-Hoyle model is in agreement with the results of \citet{Gulisano05}, where both fits were performed on individual MCs.

When we used normalised fields, the differences between the two fitted models increased (\fig{fit_B_SEA}b,d,f). The Lundquist model almost perfectly fits the SEA profiles of $\byFR$ and $\bzFR$. In contrast, the Gold-Hoyle model significantly deviates from the SEA profiles in opposite ways in the core and at the periphery of the FR. Indeed, its $\chib$ is about twice that of the Lundquist model. As above, this conclusion extends to other variants of the fits (\tab{fit}). We conclude that the Lundquist model represents closely the profile of the direction of the magnetic field derived with the SEA and the main deviation between the model and the data is due to a flatter field strength in the data (\fig{fit_B_SEA}g). This is most probably due to a flattening of the flux rope cross section in the propagating direction, as modelled with the elliptical FR of \citet{Vandas03}. 

Finally, for all cases the FR radius $r_0\approx 0.5$ with the normalised units. Next, the deduced impact parameter is similar with both models and with the methods used for the SEA (\tab{fit}). It is within the interval $[0.13,0.2]$ and with a mean of $0.17$. This is compatible with the mean of $|p|=0.15$, derived from the histogram of $\Qbest$ (\fig{histogram_p}).

\setlength{\tabcolsep}{5pt}   
\begin{table}   
\caption{Results for the SEA of field components fitted by an FR model.}
\label{tab_fit}   
\begin{tabular}{ccccccc}     
  \hline                  
model & FR frame & SEA  & $\chiB$&$100\,\chib$ & $\Bo$ &$p$ \\
      &          &      &   (nT)   &           &   (nT)  &    \\
  \hline 
  L  & Lepping   & mean &  0.77  &    3.2      & 15.8 & 0.18 \\
  L  & Lepping   &median&  0.72  &    4.0      & 15.1 & 0.15 \\
  L  &   MV      & mean &  0.80  &    2.8      & 15.5 & 0.20 \\
  L  &   MV      &median&  0.80  &    4.8      & 14.9 & 0.19 \\
  GH & Lepping   & mean &  1.20  &    6.2      & 16.8 & 0.15 \\
  GH & Lepping   &median&  1.03  &    6.4      & 16.0 & 0.13 \\
  GH &   MV      & mean &  1.23  &    6.1      & 16.6 & 0.16 \\
  GH &   MV      &median&  1.09  &    7.0      & 16.0 & 0.16 \\
  \hline
\end{tabular}
\tablefoot{The FR model corresponds to either  the Lundquist (L, \eq{B_Lundquist}) orthe  Gold-Hoyle (GH, \eq{B_GH}). The FR frame is derived from the  Lepping or MV method. The SEA is computed with the mean or the median of events. The fit parameters are described in \sect{SEA_Fitting}.}
\end{table}   

\begin{figure}[t!]           
\centering
\includegraphics[width=0.5\textwidth, clip=]{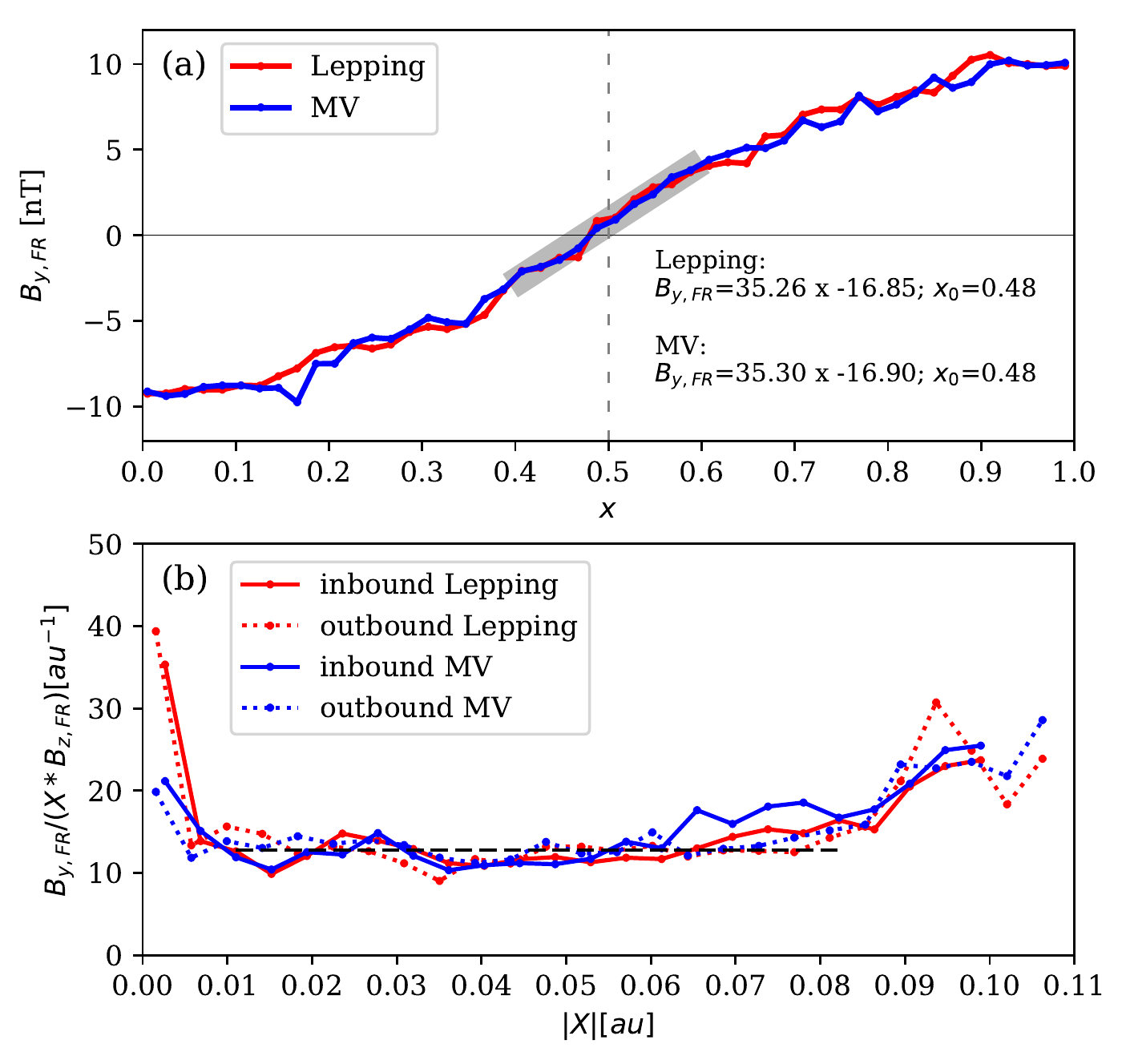}
\caption{(a) Median profile of $\ByFR$ in function of $x$. The thick grey line is the linear least square fit within the FR core for MV FR frame. The linear fit and x-intercept ($x_0$) are written for both Lepping and MV FR frame. Finally, the dashed line corresponds to  $x=0.5$. (b) $\ByFR/x\BzFR$ in function of $X$, the dashed line correspond to $\tau=12.8$~au$^{-1}$, the mean value of the twist in the range $0.01$~au~<X<~$0.08$~au considering both orientations. Inbound corresponds to $x<0.48$ and outbound to $x>0.48$ of panel (a).}
 \label{fig_twist_vs_x}
\end{figure}
 
\subsection{Generic twist profile}
\label{sect_SEA_twist}

The exact value of the twist from a given magnetic field configuration can be computed by integrating the field lines and computing the number of turns around the flux rope axis per unit length along the axis, $\tau = d\theta/dz$. This is not possible in MCs because we have only a cut close to the FR axis. We therefore need some hypothesis for simplification, such as a cylindrical symmetry. Under this hypothesis, the magnetic field twist per unit length is defined as $\tau=B_{\theta}/(r\Bz)$, with $r$ the radial distance from the FR axis.
The computation of the radius $r$ requires estimating the impact parameter $p$ because $r=\sqrt{x^2+ (p~r_0)^2}$. As the estimation of $p$ might add uncertainty in the twist computation, the field line equation for cylindrical geometry might be used, $B_{\theta}/r = B_y/x$, to avoid introducing the impact parameter (with $x=0$ set at the reversal of $B_y(x)$).

In \fig{twist_vs_x}a the median profile of $\ByFR$ is plotted for both Lepping and MV orientations in function of the dimensionless spatial coordinate $x$ as in \fig{fit_B_SEA}. The linear fit of the profiles near the FR axis are shown as a thick grey line for both orientations (Lepping and MV). The fitting results for both profiles are similar so not distinguishable on the figure. In particular, the intersection with the $x$ axis is at $x_0=0.48$ for both of them. We defined the inbound of the MC for $x<0.48$ and the outbound for $x>0.48$. We next defined the physical spatial coordinate $X$ associated with $x$ as follows. First, a shift was applied to $x$ so that $\ByFR=0$~nT corresponds to $X=0$. Second, we re-scaled the result to the MC radius. Then, we computed the physical spatial variable $X=(x-x_0)(\left \langle R_0 \right \rangle/r_0),$  with $\left \langle R_0 \right \rangle=0.12$~au being the mean value of the physical FR radius $\Ro$ reported in the Lepping list for the selected $\Qbest$ events, and $r_0=0.576$ the normalised FR radius obtained from the Lundquist model fitted to the median profile with the MV orientation done in \sect{SEA_Fitting}.
In this new variable, the inbound correspond to $X<0$ and the outbound to $X>0$. Then, the magnetic twist profile is computed as,
  \BE \label{eq_twist_vs_x}
  \tau (X) = \frac{\ByFR}{X\BzFR} \,
  \EE
The twist in function of the physical spatial coordinate $X$ is shown in \fig{twist_vs_x}b. The twist for $X>0.01$~au has a mean value of $\tau\approx12.8$~au$^{-1}$ and a slight increase toward the boundaries of the MC. In contrast, the twist presents a surprisingly strong increase while $X$ decreases to zero, a similar behaviour to the one found by \citet{Wang16}.

Next, we considered computing the twist using the azimuthal component of the magnetic field in the FR frame as $\BthetaFR = \sqrt{\BxFR^2 + \ByFR^2}$. The physical radius $R$ where the spacecraft observes the field is computed as $R=\sgn(X)\sqrt{X^2+(p\left \langle R_0 \right \rangle)^2}$ in astronomical units, with $p$ the value of the impact parameter. $p$ is estimated from fitting the Lundquist model to the median SEA profiles, reported in \tab{fit}. We found $p=0.19$ when we used the MV frame in the SEA and $p=0.15$ when we used the Lepping frame. Now, the twist expressed in function of the radius of the MC is rewritten as

  \BE \label{eq_twist_vs_r}
  \tau (R) = \frac{\BthetaFR}{R\BzFR} \,
  \EE
  
\begin{figure}[t!]           
\centering
\includegraphics[width=0.5\textwidth, clip=]{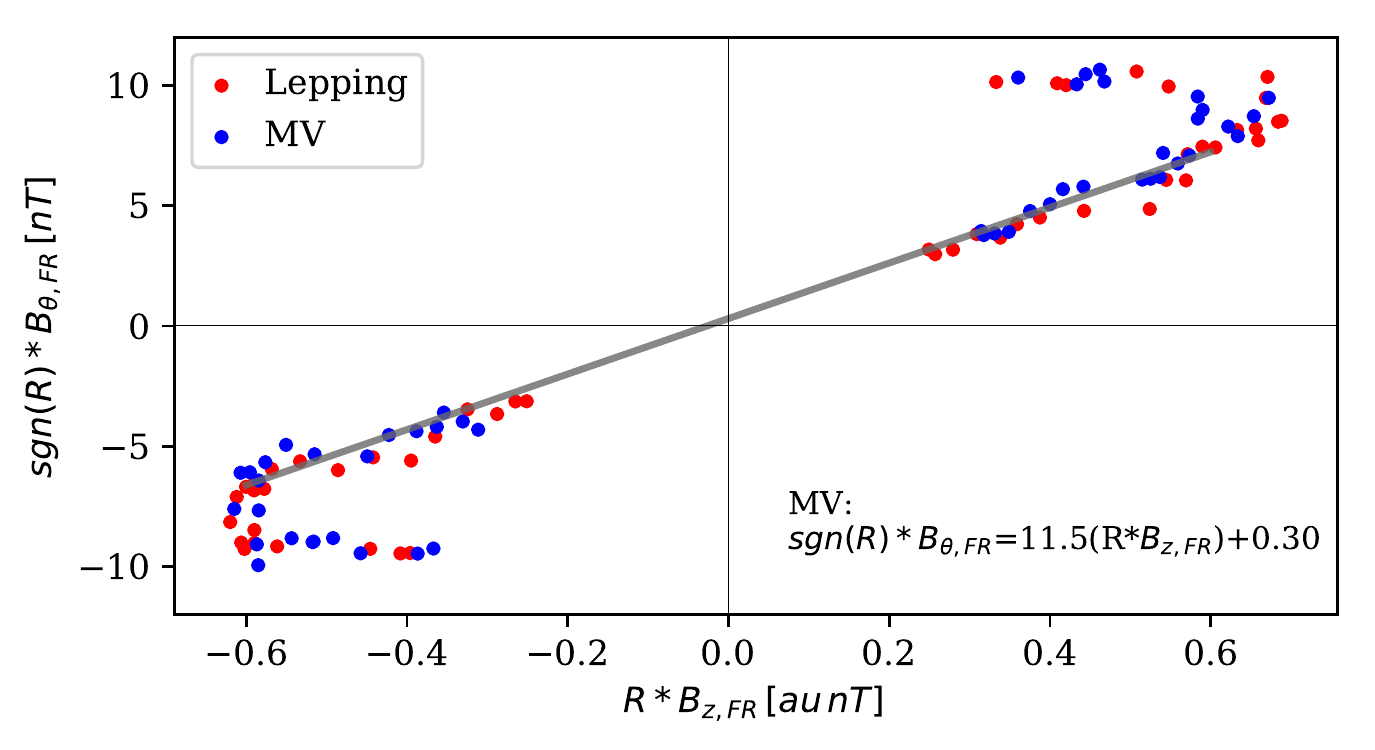}
\caption{Scatter plot of $\sgn(R) \BthetaFR$ as a function of $R\BzFR$ using the median profiles of the SEA of \fig{superposed_L_vs_MV}. The variable $R$ is considered negative in the inbound and positive in the outbound in order to separate the two regions. The grey line corresponds to the linear least square fit for the MV (blue points) in the range $-0.06$~ au$~<R<~0.06$~au. The y-intercept ($y_0=0.3$) can be approximated to zero, so that the slope of the linear fit approximates to the twist.}
 \label{fig_twist}
\end{figure}

The azimuthal component of the magnetic field in the FR frame as a function of $R\BzFR$ is shown in the \fig{twist}. In order to separate in- and outbound in the plot, we set $R<0$ for the inbound. The results from Lepping and MV present a similar behaviour.
The points are distributed along an S-shaped curve with nearly aligned points in the central bar of the S-shape.
The linear least-squares fit in this region for the MV is shown as a grey line; the slope approximates the twist in the FR core. We note that the y-intercept is not equal to zero. This shift in the fitted line can be associated with an error for the estimation in the definition of in- and outbound associated with $x_0$.

\begin{figure}[t!]           
\centering
\includegraphics[width=0.5\textwidth, clip=]{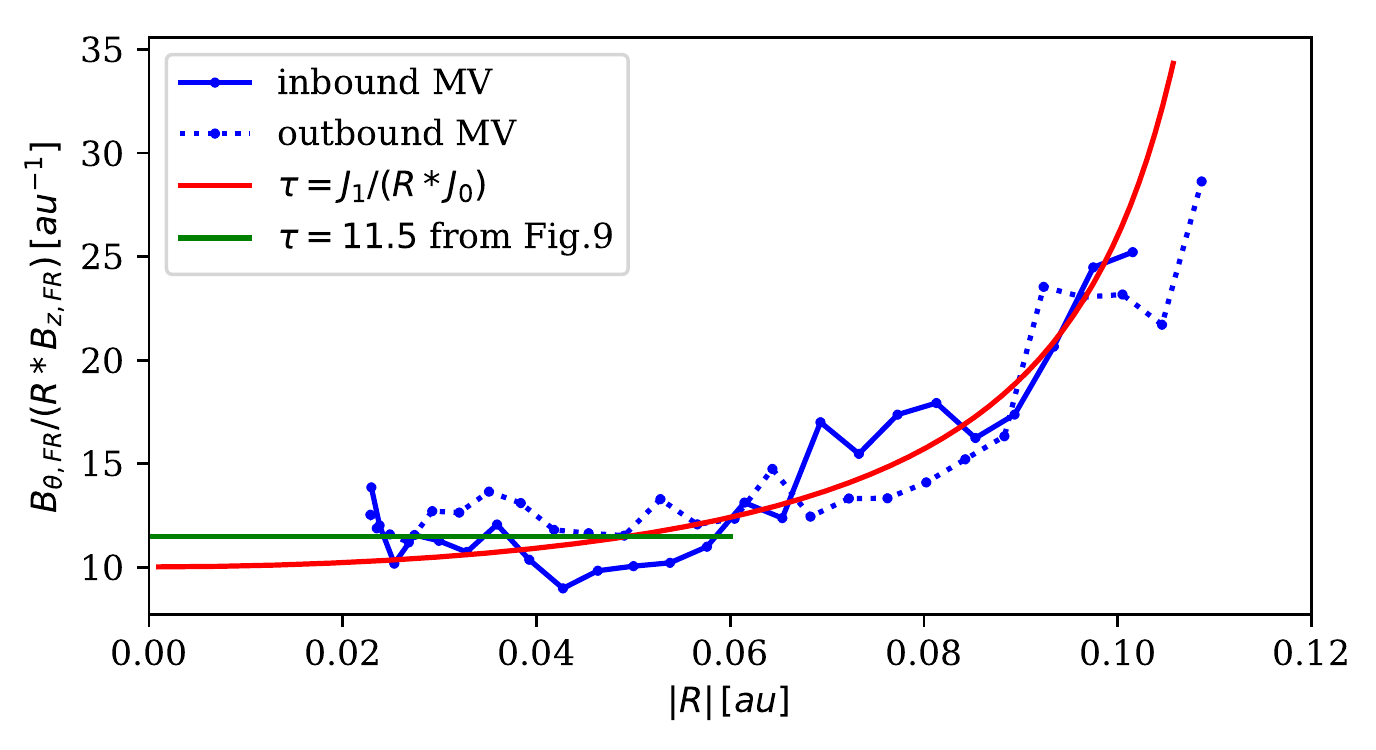}
\caption{Twist as a function of the radius $R$ computed from different methods. The blue lines (continuous and dashed) are the results derived from the median profiles of the SEA computed with FR orientation defined by the MV method. The red line is the Lundquist model fitted to the same SEA data (\fig{fit_B_SEA}), and the green line is the twist derived from the least-squares fit of a straight line in the FR core (\fig{twist}).}
 \label{fig_twist_vs_r}
\end{figure}

Finally, the twist as a function of $R$ is shown in \fig{twist_vs_r}. We present the magnetic twist profile inside the magnetic cloud with three different methods. The blue curve corresponds to the twist profile using the median profile of the SEA, with the MV orientation and an impact parameter of $p=0.19$. The red curve is obtained from the Lundquist model using a value of $\alpha=20$~au$^{-1}$ (\ie\ $R_0=0.12$~au), finally, the constant twist value derived from \fig{twist} is shown with a green line. In contrast to \fig{twist_vs_x}b , the twist is almost constant and equal to $\tau=11.5$~au$^{-1}$ for $|R|<R_0/2$ and increases moderately, up to a factor two, towards the MC boundaries. The Lundquist model presents a similar behaviour with the twist profile obtained with the SEA, although it diverges for $|R|\gtrsim R_0$. This corresponds to the right side of the figure, where no data are present ($\BzFR = 0$ for $R=R_0$ in the Lundquist model, while $\BzFR > 0$ at the MC boundary in \fig{fit_B_SEA}). Moreover, as the selection of MCs, corresponds to the most symmetric events there are no significant differences between the inbound and outbound of the MC.

As mentioned before, the magnetic twist profile shown in \fig{twist_vs_x}b presents large values of the twist for small values of $X$. In this case, the twist was computed with \eq{twist_vs_x}. Near the centre of the MC, both $\ByFR$ and $X$ approximates to zero.
This leads to a mathematical undetermination of $\tau$ and to arbitrary large values of $\tau$ in data where $X$ is approaching zero while $\ByFR$ is not null.
When we include the impact parameter and compute the magnetic twist profile in function of the radius (\eq{twist_vs_r}), the twist shows a different behaviour near the FR axis (see \fig{twist_vs_r}). In this case, $R$ decreases almost at the same rate as $\BthetaFR$ and the twist is approximately constant near the centre of the MC in agreement with the good fit of data by the Lundquist fit (\fig{fit_B_SEA}). So, we conclude that to compute the twist profile, it is better to use $R$ than $X$, then to use \eq{twist_vs_r} instead of \eq{twist_vs_x}, and that the error associated to $p$ has a weaker impact on the twist profile than the intrinsic undetermination present in the ratio of \eq{twist_vs_x}.

\section{Summary and conclusions}
\label{sect_Conclusions}

The twist of the magnetic field lines around the FR axis is an important property of MCs. 
This quantity is one of the keys to link MCs with their solar origin and ejection processes, and also to understand part of the equilibrium state (\eg\ magnetic tension). In particular, the twist is crucial to determine the content of magnetic helicity transported by MCs in the interplanetary medium. 

In situ observations can provide only a 1D cut inside the 3D magnetic structure of MCs. Observations also include an intrinsic mix between spatial structures and time evolution because a spacecraft observes different parcels of fluid at different times. Moreover, observations include different phenomena such as
(a) small-scale fluctuations, (b) dynamical sub-structures (\eg\ large amplitude Alfvén waves), or (c) local distortions from the global structure that are only present near the region observed by the spacecraft, but are not common to the full FR. 
Thus, it is very difficult to determine the twist distribution in individual MCs, even under the hypothesis of a cylindrical cross section.

The aim of this work was to estimate the typical twisted magnetic field profile inside MCs, from in situ observations at 1~au, applied to a sample of events.
We applied an SEA to a subset of the best observed MCs within the set analysed by \citet{Lepping90}.  We defined an asymmetry factor of the field strength ($\CB$) in order to exclude the most asymmetric profiles of $B(t)$ (with respect to the centre of the MC time interval).  We also set a threshold on the impact parameter, $|p|\leq 0.3$, in order to exclude cases where the FR core is not observed.  The magnetic field of each MC was rotated to the FR frame using the Lundquist method or MV orientations of the FR axis.  Then, we set all MCs cases to positive helicity and positive impact parameter, so that their components added up. Finally, the SEA was applied to the magnetic field components. This statistical tool allowed us to obtain a mean or median profile that emphasises the common characteristics of the MCs.

This procedure provides the typical magnetic field components in the FR frame.
As expected with the selection of the impact parameter, 
$\BxFR$ is positive and smaller than $B$.  
More surprising, the profile of the azimuthal component, $\ByFR$, is nearly anti-symmetric and the axial component, $\BzFR$, is nearly symmetric around the centre of the MC time interval. These results are consistent with the presence of a symmetric flux rope. 
These profiles are also much smoother than the original MC data because the SEA decreases the contribution of the individual peculiarities.

Next, we fitted the SEA results with the Lundquist and Gold-Hoyle models. Both models closely represent the profiles of the SEA; the largest deviation is obtained for the field intensity $B(t)$ with SEA profiles flatter than those of both models. Still, the Lundquist model best fits the SEA results of the B components while it is almost a perfect fit for the SEA profiles of the normalised field components (unit vectors).  This allows a precise determination of the twist profile because it is derived from the normalised components.
 Furthermore, the resulting impact parameter, $p$, obtained from the fitting procedure, is within the interval $[0.13,0.2]$ and with a mean of $0.17$.  These results are consistent with the mean value of the impact parameter of the MCs data set.

Next, we computed the twist profile, $\tau$, from three different methods. First, $\tau(X)$ was obtained from $\ByFR$ and $\BzFR$ components, with $X$ the spatial coordinate across the FR and obtained from the normalised time by assuming a constant velocity. In this case, the twist profile leads to high twist values near the FR axis because $X$ vanishes where $\ByFR$ is weak but non-zero. In the second method, the impact parameter is included, and we estimate the twist as $\tau(R)$, with $R$ the FR radius. This computation involves the azimuthal field component $\BthetaFR$. In this case, the twist is nearly constant in the FR core and increases toward the boundaries by about a factor two (\fig{twist_vs_r}). Finally, in the third method the twist profile was derived from the Lundquist model fitted to the SEA results (using the FR orientation for each event derived from the MV method before making the superposition). This provides a consistent result with the second method (\fig{twist_vs_r}). This is in contrast with the results of the first method and of \citet{Wang16} showing that the computation of a twist profile in MCs could be strongly biased, in particular in the core, if a too approximative method is used.

In summary, the results show that the Lundquist model fits the SEA results of the normalised field components very closely.  These results were confirmed by finding the orientation of FRs with MV that is independent of the Lundquist model. This result explains the broad success of the Lundquist model in fitting a large variety of MCs. Magnetic clouds typically have a twist profile as in the Lundquist model, but large variations from case to case that partly mask the common profile. 
On average, the FR core is almost uniformly twisted in about half its central part, with a twist that slightly increases by up to a factor two towards the FR boundary.  This implies that a vanishing axial field component, implying an infinite twist, should not be applied at the MC boundaries in any fitting procedure.

\begin{acknowledgements}
S.D. and V.L. acknowledge partial support from the Argentinian grants UBACyT (UBA) and PIP-CONICET-11220130100439CO.
This work was partially supported by a one-month invitation of P.D. to the Instituto de Astronom\'ia y F\'isica del Espacio and by a one-month invitation of S.D. to the Observatoire de Paris.
P.D. thank the Programme National Soleil Terre of the CNRS/INSU for financial support.
S.D. is member of the Carrera del Investigador Cien\-t\'\i fi\-co, CONICET. V.L. is fellow of CONICET. \end{acknowledgements}

%
\bibliographystyle{aa}
\bibliography{mc}
\IfFileExists{\jobname.bbl}{}
{\typeout{}
\typeout{****************************************************}
\typeout{****************************************************}
\typeout{** Please run "bibtex \jobname" to optain}
\typeout{** the bibliography and then re-run LaTeX}
\typeout{** twice to fix the references!}
\typeout{****************************************************}
\typeout{****************************************************}
\typeout{}
}

\begin{appendix}

\section{Flux ropes: frame and models}
\label{ap_FRframe}

A coordinate system that is typically used in space near Earth is the geocentric solar ecliptic (GSE), which is defined from an orthogonal base of unit vectors ($\uxGSE$, $\uyGSE$, $\uzGSE$). In this frame $\uxGSE$ and $\uyGSE$ are in the ecliptic plane, where $\uxGSE$ points from the Earth towards the Sun, $\uyGSE$ points in the direction opposite to the terrestrial rotation motion around the Sun, and $\uzGSE$ points to the north pole of the heliosphere.

\begin{figure}[t!]            
\centering
\includegraphics[width=0.5\textwidth, clip=]{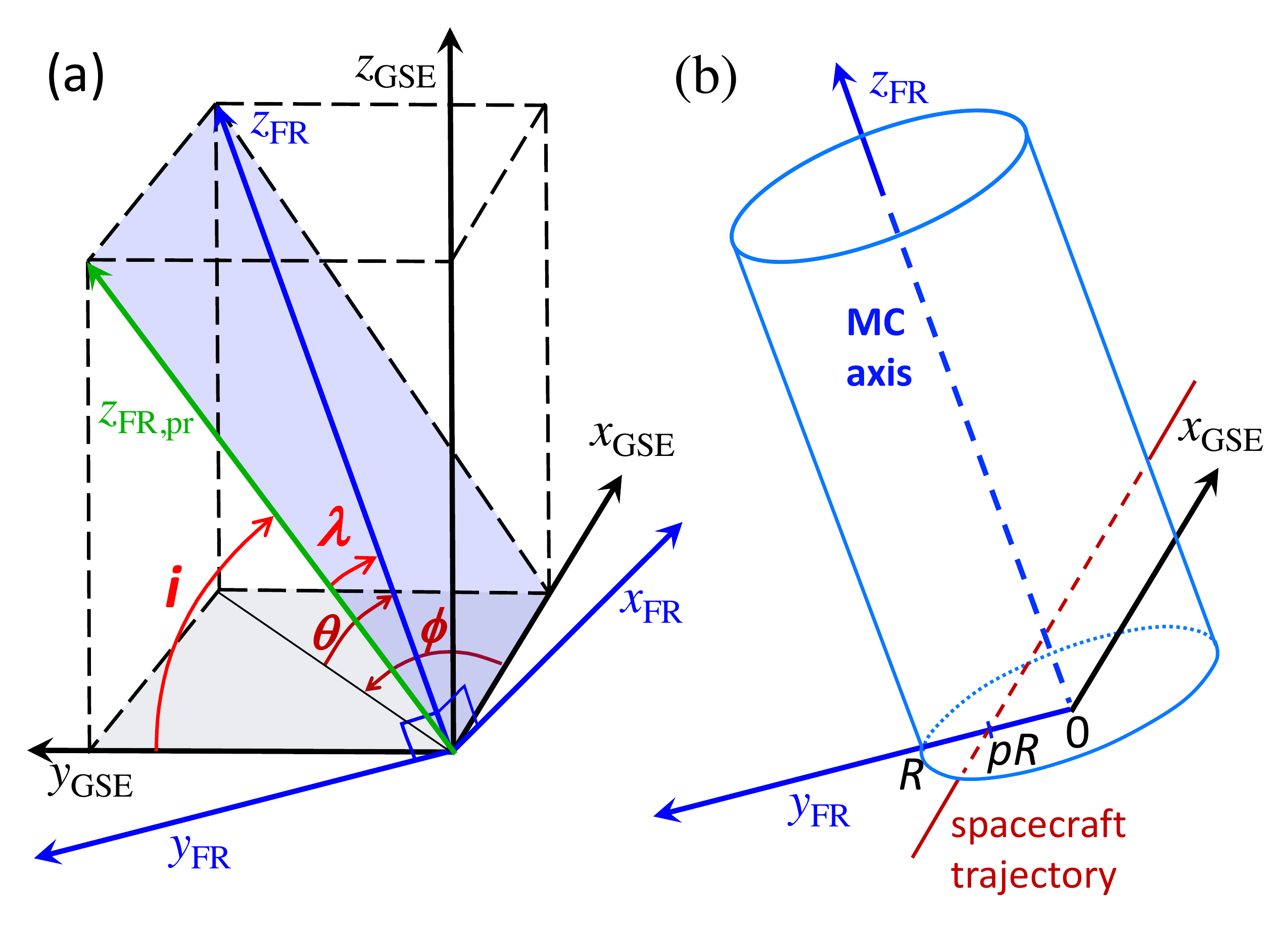}
\caption{Definitions of the GSE and FR coordinate systems, the angles defining the FR axis orientation ($\theta$, $\phi$), or ($\lA$, $\iA$), and the impact parameter $p$. In the FR frame, the rectilinear spacecraft trajectory cut $\yFR$ axis at the $p \, R$ value (with $R$ the FR radius).  
}
\label{fig_coordinate_system}
\end{figure}

The FR axis orientation is usually defined with the latitude ($\theta$) and longitude ($\phi$), considering $\uzGSE$ as the polar axis of the spherical coordinates. Thus, the latitude is the angle between the ecliptic plane and the FR axis ($\uzFR$); the longitude is defined as the angle between the projection of the FR axis onto the ecliptic plane and $\uxGSE$ in anti-clockwise direction (\fig{coordinate_system}a).

   
A non-negligible number of MCs are observed with $|\theta| \approx 90 \degree$, that is, close to the polar axis, where $\phi$ is highly variable. This implies that $\phi$ has a large error.  To solve this problem and to introduce a more physical meaning of the angles defining the FR axis, \citet{Janvier13} set the polar axis along the Sun-Earth direction ($\uxGSE$).
The FR axis of all the useful MC encounters to derive FR properties lies away from $\uxGSE$, so the new system of coordinates avoids the above problem for the useful MC encounters.  The new polar angles are $\lA$ and $\iA$
(which are new ``latitude'' and ``longitude'' with respect to this new polar axis $\uxGSE$, \fig{coordinate_system}a). 
$\lA$ is the angle between $\uzFR$ and its projection on the plane perpendicular to the Sun-Earth direction (\ie\ the plane defined by $\uyGSE$ and $\uzGSE$). And $\iA$ is defined as the angle between this projected $\uzFR$ and  $\uyGSE$ (\fig{coordinate_system}a).
$\lA$ was called the ``location angle'' since it is linked with the location where the spacecraft crosses the flux rope axis ($\lA=0$ at the flux rope apex and $|\lA|$ increases away from the apex). $i$ was called the inclination angle because it indicates the inclination of the flux rope axis respect to the ecliptic. 

The study of different properties of MCs, such as magnetic flux and twist, is best defined in the local frame, known as the FR frame.
It is defined with $\uzFR$ parallel to the axis of the flux rope, such that $\BzFR$ is positive in the FR core.  
  The vector $\uxFR$ is defined so that the supposed rectilinear spacecraft trajectory is within the plane ($\uxFR ,\uzFR$) while $\uyFR$ is orthogonal to this plane.  In the ideal case of a cylindrical FR and with a spacecraft crossing the FR axis, $\uxFR$ corresponds to the cylindrical radial direction. More generally, with a trajectory defined by the unit vector $\ud$ pointing towards the Sun, $\uyFR$ is defined in the direction $\uzFR \times \ud$ and $\uxFR$ completes the right-handed orthonormal base ($\uxFR ,\uyFR ,\uzFR $) which defines the FR frame (\fig{coordinate_system}b). The origin of the FR frame is set on the FR axis at the closest distance from the spacecraft trajectory.

The in situ plasma measurements indicate that MCs are moving at 1~au nearly radially away from the Sun at a speed much higher than that of the spacecraft.  Then, within the FR frame attached to the moving MC, the spacecraft trajectory is 
$(\xSC (t) \cos \lA, \yp, \xSC (t) \sin \lA)$. $\xSC (t)$ is the spacial coordinate along the spacecraft trajectory, and it is a function of time.  $|\yp|$ is the minimum distance of the spacecraft trajectory to the FR axis.

We next assume that in the vicinity of the spacecraft crossing, the FR has a cylindrical shape with only a radial dependence (\fig{coordinate_system}b).   In the local cylindrical coordinates ($\rc,\thc,\zc$) the magnetic field writes ($0,\Bt(\rc),\Bz(\rc)$). 
In the FR frame the azimuthal component is projected along $\uxFR $ and $\uyFR $ axis, which implies the field components:
   \BA
  \BxFR (t) &=&-\Bt(\rc) \, p  \nonumber  \\
  \ByFR (t) &=& \Bt(\rc) \, (\xSC (t) \, \cos \lA ) /R     \label{eq_B_FR} \\
  \BzFR (t) &=& \Bz (\rc)      \nonumber 
  \EA
where $p=\yp /R$ is the impact parameter, $R$ the FR radius, and
  \BE \label{eq_rc}
  \rc (t) = \sqrt{(\xSC (t) \, \cos \lA )^2 + (p\,R)^2} \,
  \EE
More complete equations, in particular with expansion, could be found in \citet{Demoulin08}.  
A classical FR example is a linear force-free field (Lundquist's FR):
  \BA
  \Bt (\rc) &=& \Bo ~J_1(\alpha \, \rc) \nonumber \\
  \Bz (\rc) &=& \Bo ~J_0(\alpha \, \rc),  \label{eq_B_Lundquist} 
  \EA
where $\Bo$ is the axial field strength of the FR axis, $\alpha$ a parameter, which defines the amount of twist around the FR axis ($=\alpha/2$), and $J_0$, $J_1$ are Bessel functions.

Another typical model used to describe the magnetic structure in MCs is the constant-twist Gold-Hoyle model, where the components are
  \BA
  \Bt (\rc) &=& \frac{\Bo \tau_0 \rc}{1+\tau_0^2 \rc^2} \nonumber \\
  \Bz (\rc) &=& \frac{\Bo}{1+\tau_0^2 \rc^2}  \label{eq_B_GH} 
  \EA
where $\tau_0$ is the constant twist of the field lines around the FR axis.
 
\onecolumn
\section{Table of selected best MCs}
\label{sect_Table}

\setlength{\tabcolsep}{9pt}
\begin{table}[h!]
\renewcommand{\arraystretch}{1.5}
    \centering
    \caption{List of the 25 events.}
    \begin{tabular}{cccccccc}
        \hline
        Event & Start &  End & $\lL$ (deg)& $\lMV$ (deg) & $\iL$ (deg)&$\iMV$ (deg) &$p$ \\ 
        \hline
         1 & 18 Oct 1995 19:48 & 20 Oct 1995 01:18 & -17 & -18 & -172 & -165 & 0.12\\
         2 & 01 Jul 1996 17:18 & 02 Jul 1996 10:18 &   3 &  9  &    6 & 4    & 0.16\\
         3 & 10 Jan 1997 05:18 & 11 Jan 1997 02:18 & 23  &  25 &  179 & 197  &-0.11\\
         4 & 22 Sep 1997 00:48 & 22 Sep 1997 17:18 & 26  &  37 &   74 & 67   & 0.03\\
         5 & 07 Nov 1997 15:48 & 08 Nov 1998 04:18 & 20  &  70 &  139 & 116  &-0.16\\
         6 & 07 Jan 1998 03:18 & 08 Jan 1998 08:18 & -31 & -37 &   73 & 71   &-0.02\\
         7 & 04 Mar 1998 14:18 & 06 Mar 1998 06:18 & 29  &  22 &   24 & 23   &-0.06\\
         8 & 24 Jun 1998 16:48 & 25 Jun 1998 21:48 & 55  &  49 &   38 & 78   &-0.23\\
         9 & 09 Aug 1999 10:48 & 10 Aug 1999 15:48 & 15  &  11 &   88 & 80   & 0.26\\
        10 & 03 Oct 2000 17:06 & 04 Oct 2000 14:06 & -26 &  -27&   37 & 35   & 0.23\\
        11 & 13 Oct 2000 18:24 & 14 Oct 2000 16:54 & 41  &  81 &  -53 & -36  & 0.11\\
        12 & 06 Nov 2000 23:06 & 07 Nov 2000 18:06 & 24  &  30 &  -10 & 8    &-0.19\\
        13 & 19 Mar 2001 23:18 & 20 Mar 2001 18:18 & 44  &  50 & -129 & -131 &-0.19\\
        14 & 22 Apr 2001 00:54 & 23 Apr 2001 01:24 & -5  &   0 & -101 & -129 & 0.05\\
        15 & 19 Mar 2002 22:54 & 20 Mar 2002 15:24 & -42 & -33 &   27 & 14   &-0.18\\
        16 & 24 Mar 2002 03:48 & 25 Mar 2002 22:48 & -15 & -12 &  144 & 156  & 0.08\\
        17 & 02 Aug 2002 07:24 & 02 Aug 2002 21:06 & 25  &  -6 & -170 & -172 & 0.11\\
        18 & 24 Jul 2004 12:48 & 25 Jul 2004 13:18 & -4  & -11 &  -21 & -40  &-0.30\\
        19 & 29 Aug 2004 18:42 & 30 Aug 2004 20:48 & -36 & -20 &  -10 & -15  & 0.06\\
        20 & 13 Apr 2006 20:36 & 14 Apr 2004 09:54 &   8 &  23 & -167 & -180 &-0.23\\
        21 & 21 May 2007 22:54 & 22 May 2004 13:36 & -55 & -68 &   62 & 67   &-0.20\\
        22 & 17 Dec 2008 03:06 & 17 Dec 2008 14:24 &  25 & 30  &  179 & 200  &-0.25\\
        23 & 10 Sep 2009 10:24 & 10 Sep 2009 16:24 & -12 & -16 &   59 & 34   &-0.01\\
        24 & 30 Mar 2011 01:30 & 31 Mar 2011 16:30 &  24 & 6   &    8 & 24   &-0.14\\
        25 & 24 Oct 2011 22:54 & 25 Oct 2011 14:36 &  16 & 10  &   42 & 50   &-0.19\\
        \hline
    \end{tabular}
    \label{tab_cases}
    \tablefoot{Events of MCs from the Lepping catalogue of quality $Q_0=1$ and $Q_0=2$ with impact parameter $|p|\leq 0.3$ and  asymmetry parameter $|\CB|\leq 0.1$.}
\end{table}
\end{appendix}

\end{document}

%% file: definitions.tex
\newcommand{\BE}{\begin{equation}}
\newcommand{\EE}{\end{equation}}
\newcommand{\BA}{\begin{eqnarray}}
\newcommand{\EA}{\end{eqnarray}}
 \newcommand{\fig}[1]{Fig.~\ref{fig_#1}}

 \newcommand{\sect}[1]{Sect.~\ref{sect_#1}}
 \newcommand{\ap}[1]{Appendix~\ref{ap_#1}}
 
 \newcommand{\eq}[1]{Eq.~(\ref{eq_#1})}
 \newcommand{\eqs}[2]{Eqs.~(\ref{eq_#1}) and (\ref{eq_#2})}

\newcommand{\eg}{{\it e.g.}}

\newcommand{\ie}{{\it i.e.}}

 \newcommand{\tab}[1]{Table~\ref{tab_#1}}

\newcommand{\degree}{\ensuremath{^\circ}}

\newcommand{\rmd}{{\rm d }}

\newcommand{\uvec}[1]{\hat{\bf #1}}

\renewcommand{\ao}{a_{0}}

\newcommand{\Bo}{B_{0}}

\newcommand{\Ba}{B_{\rm a}}
\newcommand{\Bmean}{\langle B \rangle}
\newcommand{\Bs}{B_{\rm s}}
\newcommand{\Bt}{B_{\rm \theta}}

\newcommand{\Bz}{B_{\rm z}}

\newcommand{\BxFR}{B_{\rm x,FR}}
\newcommand{\ByFR}{B_{\rm y,FR}}
\newcommand{\BzFR}{B_{\rm z,FR}}

\newcommand{\byFR}{b_{\rm y,FR}}
\newcommand{\bzFR}{b_{\rm z,FR}}
\newcommand{\BthetaFR}{B_{\rm \theta,FR}}

\newcommand{\CB}{C_{\rm B}}  
\newcommand{\chiB}{\mathlarger{\mathlarger{\chi}}_{\rm B}}  
\newcommand{\chib}{\mathlarger{\mathlarger{\chi}}_{\rm b}}

\newcommand{\ud}{\uvec{d}}

\newcommand{\kms}{\rm km~s$^{-1}$}

\newcommand{\Qall}{Q_{\rm 1,2}} 
\newcommand{\Qbest}{Q_{\rm best}}

\newcommand{\rp}{r_{\rm p}}
\newcommand{\rs}{r_{\rm s}}

\newcommand{\Ro}{R_{0}}

\newcommand{\sgn}{sgn} 
\renewcommand{\to}{t_{0}}
\newcommand{\tc}{t_{c}}
\newcommand{\tend}{t_{\rm end}} 
\newcommand{\tstart}{t_{\rm start}}

\newcommand{\Vmean}{\langle V \rangle}

\newcommand{\uxGSE}{\uvec{x}_{\rm GSE}}
\newcommand{\uyGSE}{\uvec{y}_{\rm GSE}}
\newcommand{\uzGSE}{\uvec{z}_{\rm GSE}}
\newcommand{\uxFR}{\uvec{x}_{\rm FR}}
\newcommand{\uyFR}{\uvec{y}_{\rm FR}}
\newcommand{\uzFR}{\uvec{z}_{\rm FR}}

\newcommand{\xSC}{x_{\rm SC}}
\newcommand{\xo}{x_{\rm 0}}
\newcommand{\yFR}{y_{\rm FR}}
\newcommand{\yp}{y_{\rm p}}

\newcommand{\iA}{i}  
\newcommand{\lA}{\lambda} 
\newcommand{\tL}{\theta_{\rm L}}
\newcommand{\tMV}{\theta_{\rm MV}}
\newcommand{\pL}{\phi_{\rm L}}
\newcommand{\pMV}{\phi_{\rm MV}}
\newcommand{\iL}{i_{\rm L}} 
\newcommand{\iMV}{i_{\rm MV}} 
\newcommand{\lL}{\lambda_{\rm L}} 
\newcommand{\lMV}{\lambda_{\rm MV}} 

\newcommand{\rc}{r_{c}}
\newcommand{\thc}{\theta_{\rm c}}
\newcommand{\zc}{z_{\rm c}}
  
